\documentclass[a4paper, fleqn, usenatbib, useAMS]{mnras}

\def\aap{AA}
\def\apjl{ApJL}
\def\apjs{ApJS}
\def\mnras{MNRAS}
\def\apj{ApJ}
\def\aj{AJ}
\def\pasp{PASP}
\def\nat{Nat}

\usepackage{afterpage}
\usepackage{multicol}        
\usepackage{graphicx}
\usepackage{float}
\usepackage{bm} 
\usepackage{amssymb}
\usepackage{hyperref}
\usepackage{amsmath} 
\usepackage{subfig} 
\usepackage{pdflscape}
\usepackage[export]{adjustbox}
\usepackage{mathrsfs} 
\usepackage{mathtools}

\title[The GC systems of 54 Coma UDGs]{The globular cluster systems of 54 Coma ultra-diffuse galaxies: statistical constraints from {\it HST} data}
\author[N. C. Amorisco et al.]{N. C. Amorisco$^{1, 2}$\thanks{E-mail:
nicola.amorisco@cfa.harvard.edu},  A. Monachesi$^{1,3,4}$, A. Agnello$^{5}$, S. D. M. White$^{1}$ \\
$^{1}$Max Planck Institute for Astrophysics,  Karl-Schwarzschild-Strasse 1, 85748 Garching, Germany \\
$^{2}$Institute for Theory and Computation,  Harvard-Smithsonian centre for Astrophysics,  60 Garden St.,  Cambridge,  MA 02138,  USA\\
$^{3}$Instituto de Investigacion Multidisciplinario en Ciencia y Tecnologia, Universidad de La Serena, Raul Bitran 1305, La Serena, Chile \\
$^{4}$Departamento de Fisica y Astronomia, Universidad de La Serena, Av. Juan Cisternas 1200 N, La Serena, Chile \\
$^{5}$European Southern Observatory, Karl-Schwarzschild-Strasse 2, 85748 Garching, Germany} 

\begin{document}



\maketitle

\label{firstpage}

\begin{abstract}
We use data from the {\it HST} Coma Cluster Treasury program to assess the richness of the 
Globular Cluster Systems (GCSs) of 54 Coma ultra-diffuse galaxies (UDGs), 18 of which have a half-light radius exceeding 1.5~kpc.
We use a hierarchical Bayesian method tested on a large number of mock datasets to account consistently for the high and 
spatially varying background counts in Coma. These include both background galaxies and intra-cluster GCs (ICGCs), which are 
disentangled from the population of member GCs in a probabilistic fashion. 
We find no candidate for a GCS as rich as that of the Milky Way, our sample has GCSs typical of dwarf galaxies.
For the standard relation between GCS richness and halo mass 33 galaxies have a virial mass $M_{vir}\leq10^{11}M_\odot$ 
at 90\% probability. Only three have $M_{vir}>10^{11}M_\odot$ with the same confidence. 
The mean colour and spread in colour of the UDG GCs are indistinguishable from those of the 
abundant population of ICGCs. The majority of UDGs in our sample are consistent with the 
relation between stellar mass and GC richness of `normal' dwarf galaxies. Nine systems, however, 
display GCSs that are richer by a factor of 3 or more (at 90\% probability). Six of these have sizes $\lesssim1.4$~kpc. 
Our results imply that the physical mechanisms responsible for the extended size of the UDGs and for
the enhanced GC richness of some cluster dwarfs are at most weakly correlated.

\end{abstract}

\section{Introduction}

Ultra-diffuse galaxies (UDGs) are a population of low-surface brightness systems 
(effective surface brightness $\langle\mu\rangle_r\gtrsim 24$~mag/arcsec$^2$) with stellar masses 
typical of dwarf galaxies ($7\lesssim\log M_*/M_\odot\lesssim 9$). 
Ubiquitous in nearby galaxy clusters \citep{vD15, Koda15, Mun15, vdB16, Mih15, Ve17, Le17}, UDGs have also been found
outside cluster environments \citep{MD16, RT16,IT17,Be17,Lei17,Shi17,Gr17}. In clusters, they appear as roundish featureless spheroids
\citep[e.g.][]{Ya16,Mow17}, which extend the red sequence of cluster galaxies in the colour-magnitude diagram into the regime
of dwarf galaxies \citep{Koda15,vdB16,vdB17,Gu17}, with hints of a trend to bluer colours in less dense 
environments \citep{RT16}.

\begin{figure*}
\centering
\includegraphics[width=.49\textwidth]{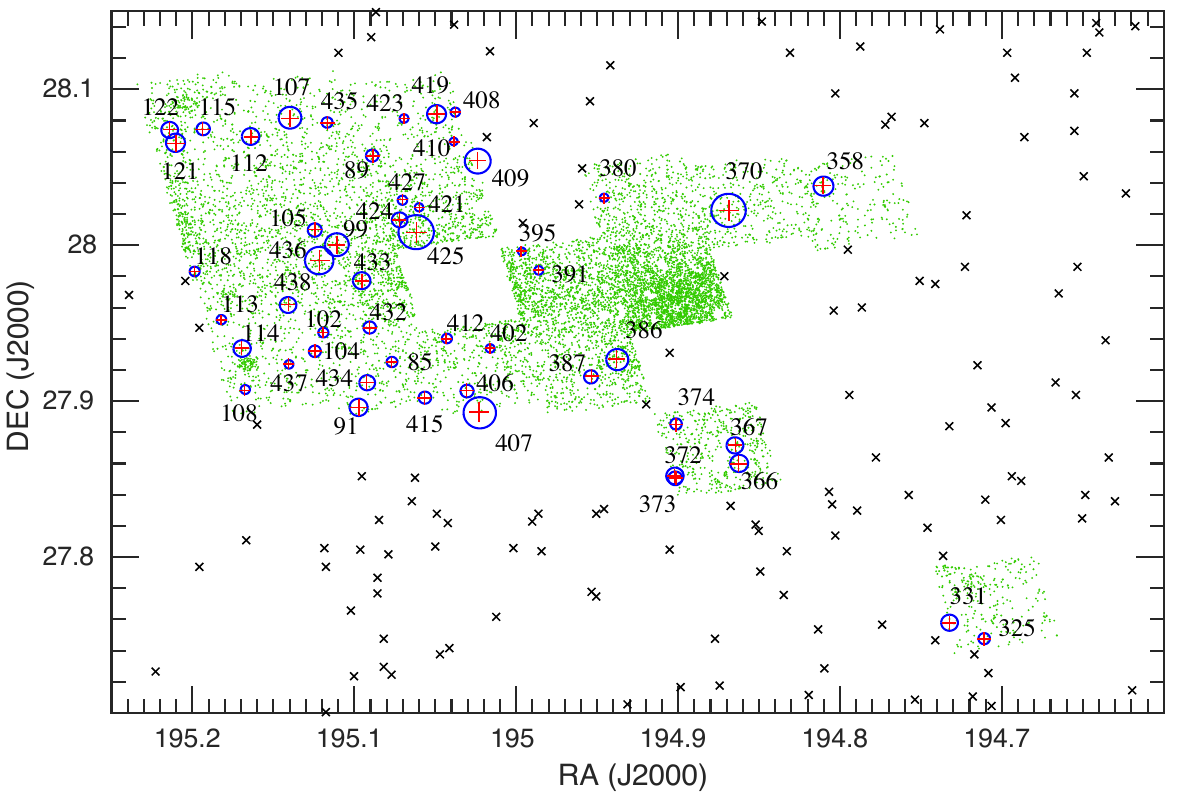}
\includegraphics[width=.49\textwidth]{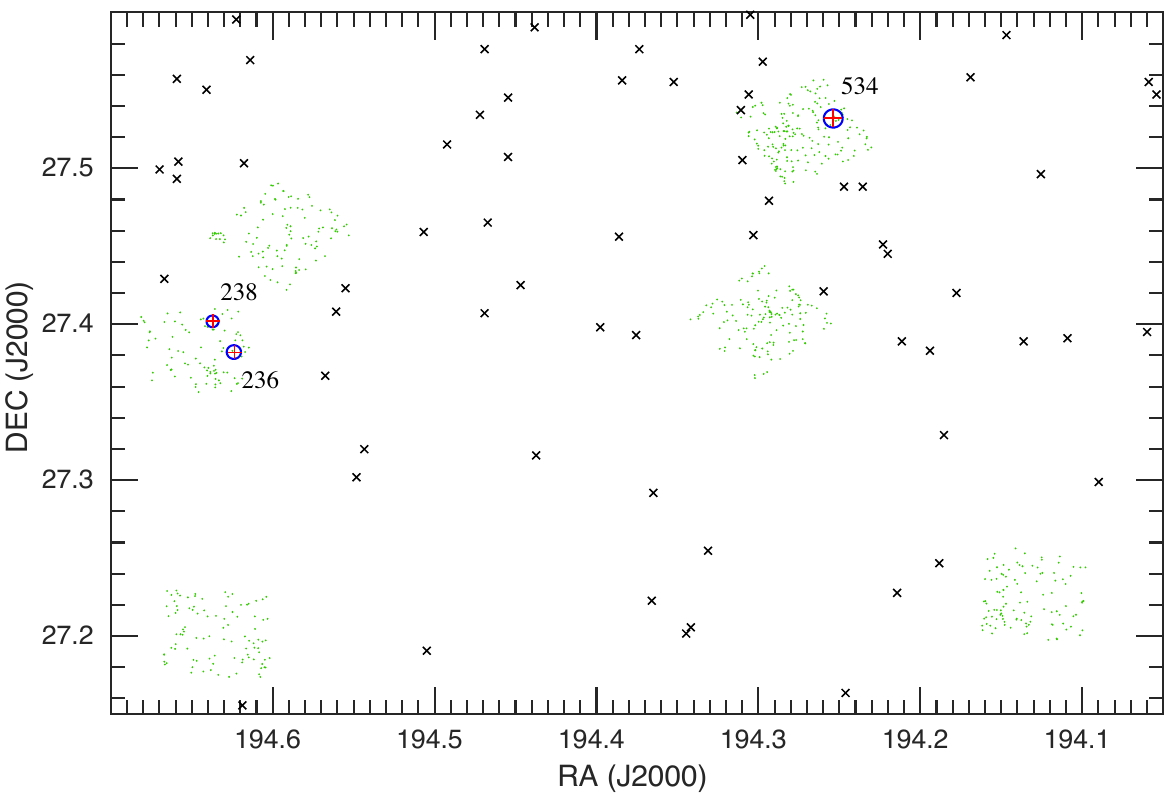}

\caption{A composite of the {\it HST}/ACS fields observed as part of the Coma Cluster Treasury program, together with the Coma UDGs from \citet{Ya16}, shown as black and red crosses. Green points are candidate GCs selected from the \citet{Ha10} catalogue, selected as in \citet{Pe11}. The 54 UDGs whose centres fall within the observed {\it HST} fields are marked in red, their size is shown by a blue circle (with a radius of $6\times R_S$), and their ID number in the \citet{Ya16} catalogue is displayed.}
\label{ACS}
\end{figure*}

UDGs are certainly highly dark matter dominated systems \citep{vD15,MB16a,NA16,vD16},
and particular interest has been sparked by the mismatch between 
their luminosity and their sizes, prompting the proposal that their halo mass 
could be much larger than suggested by their stellar mass \citep{vD15,Koda15,vD16}. 
Within this scenario, UDGs are prematurely quenched galaxies, which `fail' to 
form their stars because of their early infall onto the galaxy cluster \citep[][]{YB15,vD15}.
The largest UDGs (e.g. half-light radii $\gtrsim$1.5~kpc)
could then be hosted by Milky Way (MW) mass haloes rather than by haloes with masses below or similar to 
that of the Large Magellanic Cloud (LMC). 

One clue to the nature of UDGs 
is the almost perfect linearity of the relation between the abundance of UDGs in clusters and the 
cluster mass itself \citep{vdB16,vdB17,Ja17}. Within the LCDM framework, a linear relation is expected
if the formation mechanism of UDGs is independent of environment. As the shape of the subhalo mass function is independent
of the virial mass of the central \citep[e.g.,][]{Gao04,CG08}, a linear relation emerges naturally if UDGs are just a fraction
of the cluster subhalo population with the relevant mass.
This suggests that the physical mechanism that gives UDGs their unusual properties has an `internal' origin, and,
in contrast with the scenario above, that it is unrelated to the interaction with the cluster environment. 
This is corroborated by the detection of UDGs 
outside cluster cores \citep[][]{MD16, RT16} and in the field \citep{IT17, Be17}. 

If hosted by MW mass haloes, UDGs would lie far from the standard $M_*-M_{\rm vir}$ relation, requiring their formation 
pathway to differ fundamentally from that of `normal' haloes of the same total mass. In general, haloes of MW 
mass appear to be the most efficient at converting gas into stars \citep[e.g.][and references therein]{Guo10, PB13,BM13} 
so the UDG haloes would have to be exceptional objects with very low efficiency.
If on the other hand UDG haloes are similar in mass to those of normal dwarf galaxies, their properties could be 
accommodated by  a simple $\Lambda$CDM framework in which
they are just the low surface brightness tail of the abundant population of dwarf galaxies \citep{NA16}. This picture is 
consistent with UDGs existing both inside and outside clusters and needs no {\it ad hoc} mechanism to make them 
depart from the $M_*-M_{\rm vir}$ relation. Additionally, if hosted by low mass haloes, stellar feedback during their 
formation might lead to expansion and so contribute to their large sizes \citep[][]{AD16}.

Unfortunately, only a handful of mass measurements are available so far for UDGs.
Apart from the recent stacked analysis of \citet{Sif17}, these are all indirect, based either 
on the richness of the globular cluster system (GCS), 
or on extrapolation to the virial radius of a dynamical mass estimated in the main stellar
body of the galaxy \citep{MB16a, MB16b, Pe16, vD16}. Both techniques have their limitations. 
The approximate linearity of the relation between GCS richness and halo virial mass 
is supported by a solid pool of evidence, at least for `normal' galaxies \citep[e.g.,][and 
references therein]{Ha13,Hu14,Fo16}, but the 
mean conversion factor remains uncertain \citep[][hereafter G10]{Ha15,Za16,Ha17,IG10}. 
Dynamical measurements, however, do not guarantee higher precision, as {they
can only estimate the mass enclosed in the central regions of the galaxy, where dynamical tracers 
(stars or GCs) are present. The extrapolation from the galaxy's half-light radius to the virial radius is very substantial} \citep[e.g.][]{Wa09,Wo10,NA11,Ca16}. 

In this paper, we increase the number of virial mass estimates for UDGs by  
a factor $>3$ using imaging data from the {\it HST} Coma Cluster Treasury program to constrain the richness of the GCS of 54 Coma UDGs. 
In comparison to ground-based data, the high resolution of {\it HST}/ACS data helps significantly in
distinguishing  candidate GCs from background galaxies \citep[e.g.,][]{Pe11,MB16b,Pe16}. 
However, the high contamination rate by the intracluster population of GCs in Coma and by the 
abundant population of background galaxies implies that careful
statistical analysis is needed to gather reliable constraints.
Section~2 describes the data we use for this analysis. Section~3 sets out our hierarchical Bayesian approach, which is tested in Appendix A. 
Section~4 presents our results, which are discussed in Section~5, where conclusions are laid out.

\section{Observations and Methods}

We use the compilation of 854 Coma low surface brightness galaxies presented by \citet[][hereafter Y16]{Ya16}, based on {Subaru} Suprime-Cam 
archival data analysed in \citet{Koda15}. These are selected to have $\langle\mu\rangle_R> 24$~mag/arcsec$^2$ and 
a stellar half light radius $\geq0.7$~kpc. Among these, we select those systems whose centres lie within the footprint
of the Coma Cluster Treasury program, which we use to explore the properties of their GCSs. There are 54 such galaxies,
including 18 with stellar half light radius $\geq1.5$~kpc, which is the criterion used to define UDGs by \citet{vD15}.  
In the following, however, we refer to all of our 54 low surface brightness galaxies as UDGs. 
Their locations are displayed in Fig.~\ref{ACS}, together with their ID numbers in the Y16 catalogue.
We adopt half-light radii $R_S$ from the single Sersic fits presented by Y16. Where these were 
not deemed reliable, for example because of light from nearby systems, we adopt the listed values returned by SExtractor.

\subsection{Candidate GCs}

We cross-correlate the position of the Y16 UDGs with the catalogue of the HST/ACS Coma Cluster Treasury program (CCTp)
presented by \citet[][hereafter H10]{Ha10}. This lists all SExtractor sources detected in the deep $F814W$
images, as well as the measurements for the $F475W$ images. 
The $F814W-$band photometry is 80\% and 50\% complete at 26.8 and 27.3, respectively 
\citep[H10 and][]{Pe11}. This defines our completeness function $S_{814}$, for which we adopt the functional form suggested by
\citet[][eqn.~3, resulting in $\alpha=1.5$]{Sa15}. Assuming UDGs have a dwarf-like GC luminosity function (GCLF), 
CCTp data is $\sim$50\% {complete at the turnover, $F814W=27.33$~mag, \citep{MB16b}. Additionally, if the spread of 
the GCLF is also in line with that of dwarf galaxies \citep[Gaussian spread of 1.1~mag, G10,][]{Mi07,Pe09}}, statistically, 
49\% of all member GCs are indeed detected in the CCTp data.	

Given its pixel-size ($0.\arcsec 05/\text{pixel}$) {\it HST}/ACS imaging is well suited to disentangle 
GC candidates (GCCs) 
-- which appear as point sources at the distance of Coma -- from background galaxies, 
most of which are resolved. The H10 catalogue flags `point sources' (\verb+FLAGS_OBJ=1+),
based on photometry at different apertures. However, we find this flag to be unreliable for sources 
that are close to the UDGs' centres as a consequence of the fact that the galaxy light had not been 
subtracted prior to the production of the catalogue itself.
Because of the UDGs' contribution to the aperture flux, compact objects close to 
the centre of some UDGs may appear extended and be classified as such in the H10 catalogue. We therefore 
do not consider this flag and correct the aperture photometry of all sources in the catalogue. 
We do so explicitly on a source by source basis, by subtracting the flux contributed by each UDG
at the source's location, within the considered aperture. For this purpose, we use the UDG Sersic surface brightness 
profiles measured by Y16 \citep[we take $I_{814}\approx R$ for systems on the red sequence, H10,][]{Koda15}. 
We perform an analogous correction to the $F475W$ 
measurements, assuming that all UDGs have the same colour \citep[$F475W-F814W\approx0.85$,][]{Koda15}.

Figure~\ref{locus} shows the H10 catalogue sources in the plane of the $F814W_{1.2}-F814W_4$ concentration index
against the $F814W_4$ photometry ($F814W_{1.2}$ and $F814W_{4}$ are respectively the 1.2-pixel radius and 4-pixel  
aperture radius photometries) corrected for the above-mentioned UDG flux contribution. The vertical plume of point 
sources is clearly visible at $F814W_{1.2}-F814W_4\approx 1.0$, mainly composed of 
the abundant population of Coma intracluster GCs \citep[ICGCs,][]{Pe11}. Red arrows in the same Figure show the 
effect of the flux correction for individual sources (arrows extend between the properties 
of each of these sources before and after correction). Only the arrows corresponding to sources that satisfy 
$0.5<F814W_{1.2}-F814W_4<1.5$ after correction but not before correction are displayed, showing that the 
contribution of the UDGs' surface brightness may indeed cause some sources in the H10 catalogue to appear 
unduly extended. This effect is important
only for those source that lie close to the UDG centre, at radii $\gtrsim2 R_S$ sources are unaffected. However, 
this effect cannot be neglected, despite the low surface brightness nature of the UDGs' contribution to the aperture flux. 
As a consequence, we refrain from selecting GCCs through sharp cuts on the catalogue in either concentration 
index or in colour. {There are a number of reasons why this strategy of accounting for the background 
flux may not be ideal. These include: i) PSF blur in the ground based Y16 data; ii) the possible bias introduced by any central 
compact nuclei in the Sersic parameters; iii) the possible presence of neighboring objects that would be masked
out in the fitting in Y16. As a result, the accuracy of the corrected aperture photometry of the individual sources may be compromised, 
but, as we explain in the following section, our analysis does not use these values directly. In fact, a
posteriori, we find that this method of subtracting the background is effective and well suited for our purpose. 
The identification of abundant populations concentrated around the UDG centres gives credence to the adopted 
estimate of the background. 
These are found to have concentration indexes that are 
statistically the same as those of the point sources at large galactocentric
distances, which are unaffected by either UDG light or our implementation of the background subtraction.
{As we will show, our analysis identifies at high confidence a number of systems with 
rich GCSs. In their central regions, the surface density of sources recognized as member GCs is exceedingly
high for any significant fraction of them to be contributed by misclassified extended galaxies. 
If this was the case, we would detect systematically lower surface densities of extended sources 
in the rich UDGs. We have made this check and found no evidence for it, which gives additional credence to
our background subtraction.}

In the following, and as explained in Section~3.1 below, we disentangle background galaxies 
from GCs in Coma by explicitly modelling both 
concentration and colour distributions, by taking into account measurement uncertainties on a source by source basis. 
We only remove those sources with $F814W_4<22$, to avoid foreground MW stars and saturated pixels (H10).

\begin{figure}
\centering
\includegraphics[width=.99\columnwidth]{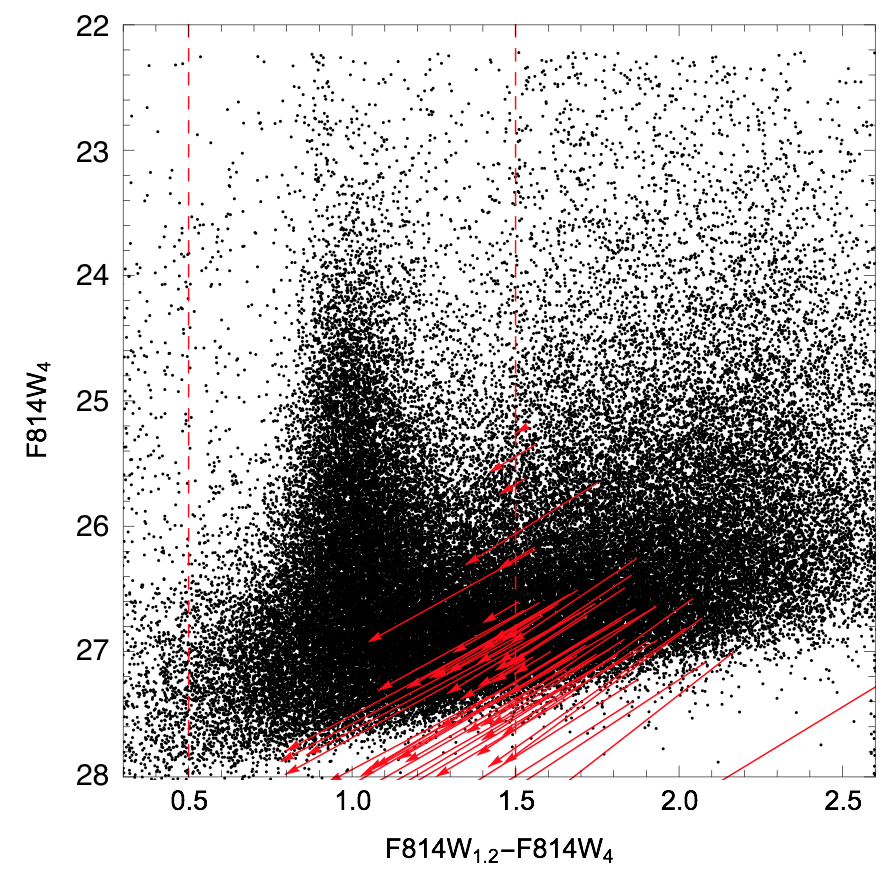}
\caption{Aperture photometry of sources in the H10 catalogue. The plume of point sources is 
clearly evident at  $F814W_{1.2}-F814W_4\approx 1.0$. Red arrows show those catalogue sources that
satisfy $0.5<F814W_{1.2}-F814W_4<1.5$ only after correcting for the UDGs' contribution to the flux
within the relevant apertures.}
\label{locus}
\end{figure}

\section{Statistical analysis}

As shown by \citet{Pe11} using these same data and confirmed by Fig.~\ref{locus}, the Coma cluster 
possesses an abundant population of ICGCs. In addition, many tens of background galaxies
are detected in each ACS field.
As is readily seen by eye \citep[see our Fig.~1 and Fig.~3 in][]{Pe11}, the CCTp distribution of GCCs 
displays clear overdensities at the locations of high surface brightness Coma galaxies. In contrast, and despite our 
above estimate of the depth of the CCTp data, visual inspection does not always reveal a concentration of GCCs near the UDG centres.
As our analysis will show, the combination of ICGCs and background galaxies is often dominant over the UDGs' GCSs (see Section~4).
As a result, we cannot reliably measure their GC 
abundances simply by counting catalogue sources within some centred aperture 
and then subtracting an estimate for the background contamination.
A more detailed statistical approach, which we describe in the following, is necessary to 
constrain the richness, $N_{{\rm GC}}$ of the GCSs of our UDG sample.

\subsection{The mixture models}

We isolate all H10 sources in the vicinity of each UDG, and model them as the superposition of 
a slowly spatially varying population of contaminants 
and a centred population of GCs physically associated with the UDG, often referred to as `members' in the following. 
We need to include two different contaminant populations, to include both resolved sources (background galaxies) and 
contaminants that appear as point sources. {The latter population is composed of ICGCs and any 
remaining foreground stars and unresolved background galaxies that appear as point sources. However, as we show in 
Section~4, our analysis suggests that the contribution of foreground stars and unresolved background galaxies is negligible, 
and that this population is indeed dominated by ICGCs. } 
In the absence of close luminous galaxies, 
we use all catalogue sources that lie within $35\times R_S$ from the UDG's centre. This is a compromise between getting 
better statistics for the contaminants and modelling their spatial distribution as locally uniform, 
with surface densities $\Sigma_{ICGCs}$ and $\Sigma_{\rm gal}$. When near luminous galaxies, 
we reduce this region on a case by case basis in order to minimise spatially variable 
contamination (see below and Fig.~\ref{mosaic}). For convenience, we centre all sources' coordinates on the UDG centre, 
and rescale them by the stellar half-light radius $R_S$. The spatial distribution of the UDG GCS is modelled with 
a Plummer profile \citep{Pl11} 
\begin{equation}
\Sigma(R)={1\over\pi} {1\over{R_h^2 \left(1+R^2/R_h^2\right)^2}} \ ,
\end{equation} 
in which {$R$ is the projected galactocentric radius} and $R_{\rm h}/R_S$, the ratio between the half-number radius $R_{\rm h}$ 
and the stellar half-light radius $R_S$, is a free parameter, different for each UDG. 
Experiments using an exponential density profile $\Sigma$ show that the profile shape does 
not affect our results (see our suite of tests in Appendix A).

As mentioned in Section~2.1 we use both colour
\begin{equation}
\textsc{c}\equiv F475W_{1.2}-F814W_{1.2} \ ,
\end{equation} 
and concentration index
\begin{equation}
C_{1.2-4}\equiv F814W_{1.2}-F814W_{4} \ ,
\end{equation} 
to disentangle member GCs from the two contaminant populations. 
As both member GCs and ICGCs appear as point sources, we assume that both populations have
the same distribution in concentration index, which we take to be Gaussian in shape. The mean and 
dispersion of this Gaussian, $\langle C_{1.2-4}\rangle_{GC}$ and $\sigma_{GC}(C_{1.2-4})$
are free parameters of the model and are fitted for using the corrected H10 catalogue. 
A different Gaussian probability distribution is used to model the distribution of concentration index of the 
extended contaminants, with mean $\langle C_{1.2-4}\rangle_{\rm gal}$ and dispersion $\sigma_{\rm gal}(C_{1.2-4})$.
These 4 free parameters are in fact hyper-parameters: they are shared by all of the 54 UDGs in our sample
and constraints are determined accordingly, using a hierarchical Bayesian approach. 
The same is true for the properties of the colour distribution of the extended contaminants, which, for simplicity, 
we also describe with a Gaussian function, with $\langle \textsc{c}\rangle_{\rm gal}$ and dispersion $\sigma_{\rm gal}(\textsc{c})$.
Instead, the population of ICGCs and the population of member GCs are allowed to have different colour distributions (in both mean and dispersion)
on a UDG by UDG basis, so to explore the properties of the UDG GCSs, and because
ICGCs may vary in different fields, for example as a function of the UDG's distance from the centre of Coma. 
Of course, for both colour and concentration index, we do take into account measurement uncertainties 
on a source by source basis. As we show in Appendix A, use of
the magnitude distribution of the sources does not improve the separation of contaminants and member GCs, 
or the accuracy and precision of the inference on the properties of the UDG GCSs. Given this, we do not use 
the magnitude distribution in the following analysis.

In conclusion, for the UDG $j$ in our sample, the likelihood of our 3 component model is \citep[e.g.,][]{WP11,NA14}
\begin{equation}
\mathcal{L}_j = \prod_i^{N_j} \left(p_{GC,i} + p_{ICGC,i} +  p_{{\rm gal},i}\right)\ ,
\end{equation} 
where $i$ runs on the H10 sources in the field defined by the UDG $j$, and, for each source,
$p_{X,i}$ is the probability of membership in the population $X$. More explicitly,
\begin{equation}
p_{GC,i}= f_{GC,j}{{S_{{\rm sp},j}({\bf r_i})\ \Sigma_j(r_i)}\over{\int S_{{\rm sp},j}\Sigma_j(r)}}\ \mathcal{G}_{GC}(C_i)\ \mathcal{G}_{GCs,j}(\textsc{c}_i)\ ,
\end{equation} 
\begin{equation}
p_{ICGC,i}= f_{ICGC,j}{{S_{{\rm sp},j}({\bf r_i})}\over{\int S_{{\rm sp},j}}}\ \mathcal{G}_{GC}(C_i)\ \mathcal{G}_{ICGCs,j}(\textsc{c}_i)\ ,
\end{equation} 
\begin{equation}
p_{gal,i}= (1-f_{GC,j}-f_{ICGC,j}){{S_{{\rm sp},j}({\bf r_i})}\over{\int S_{{\rm sp},j}}}\ \mathcal{G}_{\rm gal}(C_i)\ \mathcal{G}_{\rm gal}(\textsc{c}_i)\ .
\end{equation} 
Here, 
\begin{itemize}
\item{$f_{GC,j}$ is the fraction of the total number $N_j$ of GCCs in the studied area which are members of the GCS of the UDG $j$. 
Analogously, the fraction $f_{ICGC,j}$ are ICGCs in the relative field, while the remainder $1-f_{GC,j}-f_{ICGC,j}$ are extended contaminants.}
\item{$S_{{\rm sp},j}({\bf r})$ is the spatial selection function associated with the UDG $j$, whose value is either 0 or 1. This function 
accounts for the fact that the area available to study may be limited by the edges of the footprint, 
or by excised regions surrounding luminous galaxies, in which case its value is 0. The associated spatial integrals extend out to 35 $R_S$.}
\item{$C_i$ and $\textsc{c}_i$ are the concentration index and colour of the source $i$ and $\mathcal{G}$ is their gaussian distribution. 
The distribution of the concentration index $\mathcal{G}_{GC}(C)$ is common to all UDGs (and to both GCs and ICGCs), as are $\mathcal{G}_{\rm gal}(C)$ and $\mathcal{G}_{GC}(\textsc{c})$. The colour distributions $\mathcal{G}_{GC,j}(\textsc{c})$ and $\mathcal{G}_{ICGC,j}(\textsc{c})$ have different mean and spread for each UDG.}
\end{itemize} 

In summary, the likelihood $\mathcal{L}_j$ involves a total of 7 free parameters per UDG: the dimensionless fractions $f_{GC,j}$ and $f_{ICGC,j}$,
the dimensionless ratio $R_{\rm h}/R_S$ between the characteristic radius of the GCS and the UDG's effective radius, mean and spread
of the colour distributions of member GCs and ICGCs. In addition, the model includes 6 hyper-parameters: mean and spread
of the concentration index of point- and extended- sources, mean and spread of the colour distribution of the extended sources.
Inference on the latter set is obtained adopting a hierarchical approach, using the model likelihood
\begin{equation}
\mathcal{L} = \prod_j^{54} \mathcal{L}_j\ ,
\end{equation} 
which employs data from the entire sample of 54 UDGs at the same time.

Following the tests presented in Appendix A our priors are defined as:
\begin{itemize}
\item{{uniform in $\log f$ for both $f_{GC}$ and $f_{ICGC}$, with the constraint that $f_{GC}+f_{ICGC}<1$; in addition, we require
$\log f_{GC}>\log f_0 -1.5$, where the value of $f_0$ is estimated based on the expectation that the UDG has a `normal' GC abundance
for its stellar mass (see Appendix A2 for details), and $\log f_{ICGC} > \log f_1 - 1.5$, where $f_1$ corresponds to a total of 1 member of
the ICGC population in the whole field};  }
\item{uniform in $\log R_{\rm h}/R_S$, in the interval $0.75<R_{\rm h}/R_S<3.5$;}
\item{uniform in $\langle \textsc{c}\rangle$ {in the interval $0<\langle \textsc{c}\rangle<3$} for all of $\langle \textsc{c}\rangle_{GC}$, 
$\langle \textsc{c}\rangle_{ICGC}$, $\langle \textsc{c}\rangle_{\rm gal}$;}
\item{uniform in $\log \sigma(\textsc{c})$ for all model populations, {with $-1.75<\log \sigma(\textsc{c})<0$ };}
\item{uniform in $\langle C_{1.2-4}\rangle$ {in the interval $0<\langle C_{1.2-4}\rangle<2$ for all model populations}; 
uniform in $\log \sigma(C_{1.2-4})$ {in the interval $-3<\log \sigma(C_{1.2-4})<-0.5$ for GCs and ICGCs and 
$-3<\log \sigma(C_{1.2-4})<0.5$ for the extended contaminants}.}
\end{itemize}

\subsection{Completeness correction}

The mixture model just described allows us to infer the joint posterior distribution of the two dimensionless 
free parameters $f_{GC}$ and $R_{\rm h}/R_S$ which characterise each GCS. 
For each UDG, our final inference on the total abundance of the GCS, $N_{GC}$, is obtained by taking into account both  
spatial and magnitude incompleteness: 
\begin{equation}
N^l_{{\rm GC}}=N f^l_{GC}\times  {{\int \Sigma(r,R_{\rm h}^l)}\over{\int S_{{\rm sp}}({\bf r})\Sigma(r,R_{\rm h}^l)}}{{\int g_{GC}(F814W)}\over{\int S_{814}g_{GC}(F814W)}} \ ,
\label{incompl}
\end{equation}
where $N$ is the number of H10 sources in the UDG field, the index $l$ runs over our Markov chains, 
$S_{814}$ is the completeness function defined in Section~2.1 and 
$g_{GC}(F814W)$ is the GCLF of the member GCs. Since we will not be able to fully characterise $g_{GC}$ 
and because of the additional uncertainties introduced by the flux correction described in Section~2.1, 
we adopt a fixed Gaussian GCLF with parameters typical for dwarf galaxies: 
a mean of $\langle F814W\rangle=27.33$~mag,
and a spread of $\sigma_{F814W}=1.1$~mag \citep[G10,][]{Mi07,Pe09}. This implies a correction of a factor $\approx 2$.
It is worth noting that the turnover and spread of the GCLF 
become respectively fainter and tighter in dwarf galaxies \citep[e.g.][]{Jo07}. Therefore, our assumptions on the 
properties of $g_{GC}$ are conservative: using the GCLF of a bright galaxy would imply a smaller completeness correction, 
and therefore a lower GC richness for the same inferred value of $f_{GC}$. {Over the entire sample of 54 UDGs,
our analysis uses a total of 42703 distinct catalogue sources. }

\begin{landscape}
\begin{figure}
\centering
\includegraphics[width=\columnwidth]{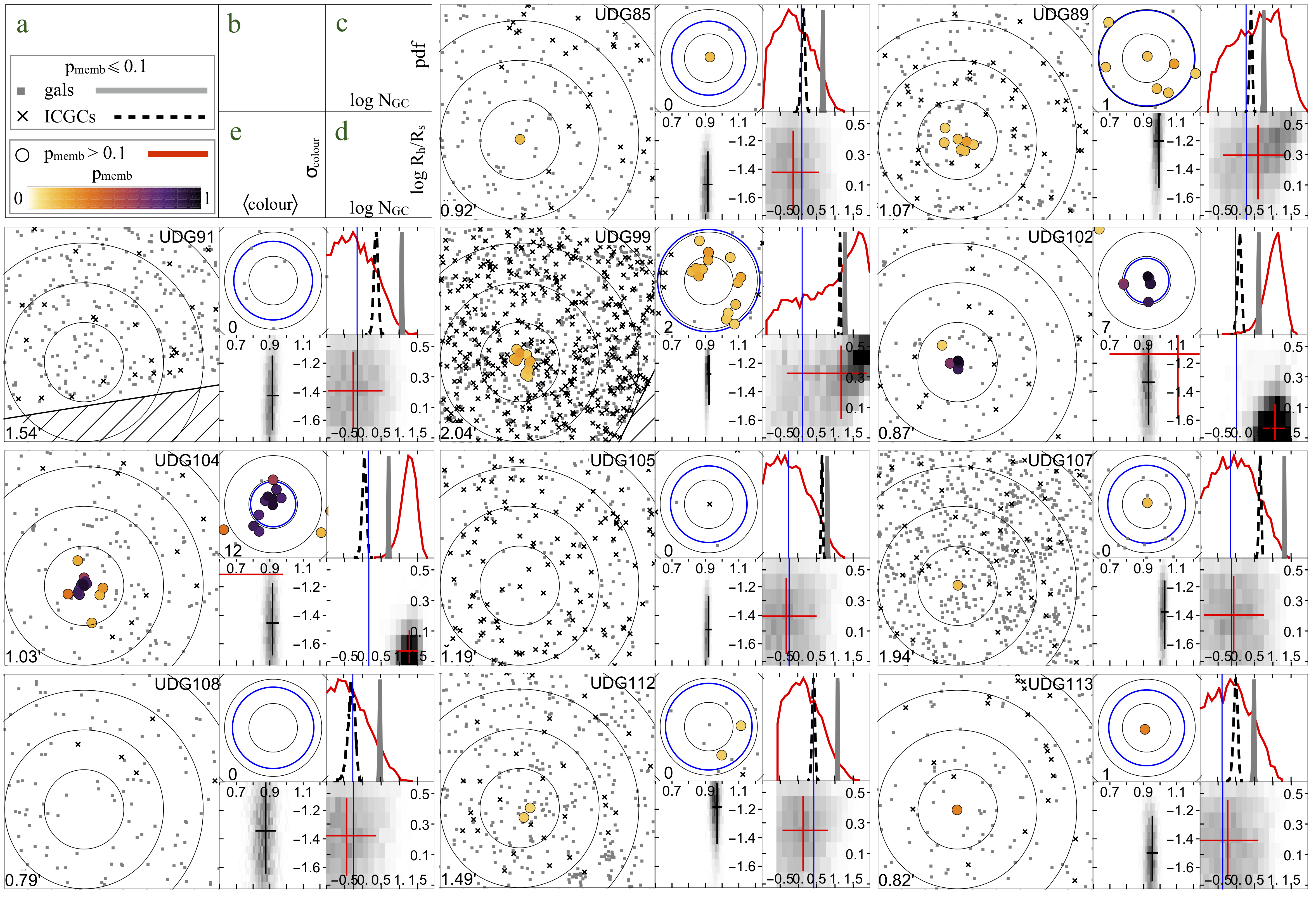}
\end{figure}
\end{landscape}
\begin{landscape}
\begin{figure}
\centering
\includegraphics[width=\columnwidth]{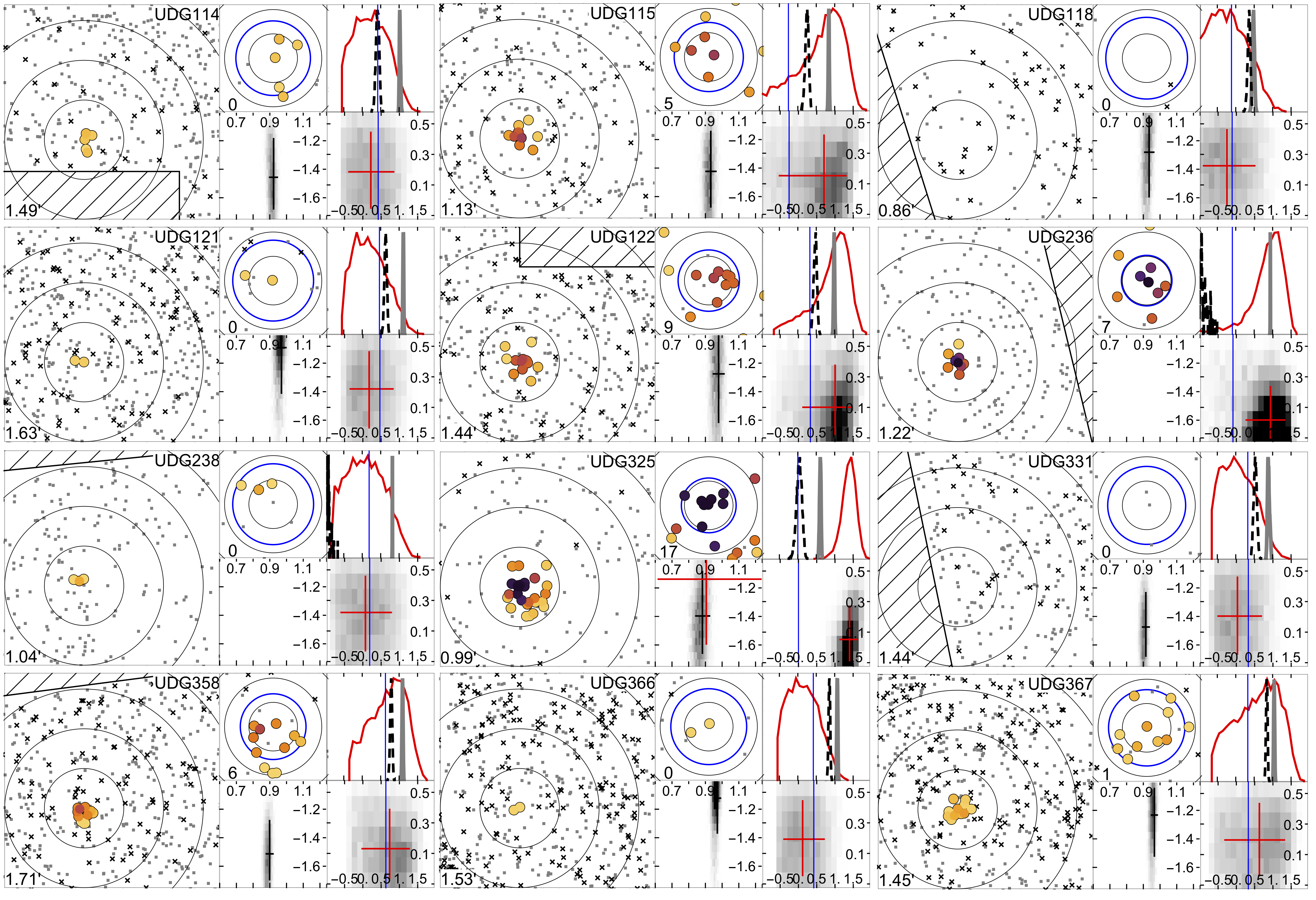}
\end{figure}
\end{landscape}
\begin{landscape}
\begin{figure}
\centering
\includegraphics[width=\columnwidth]{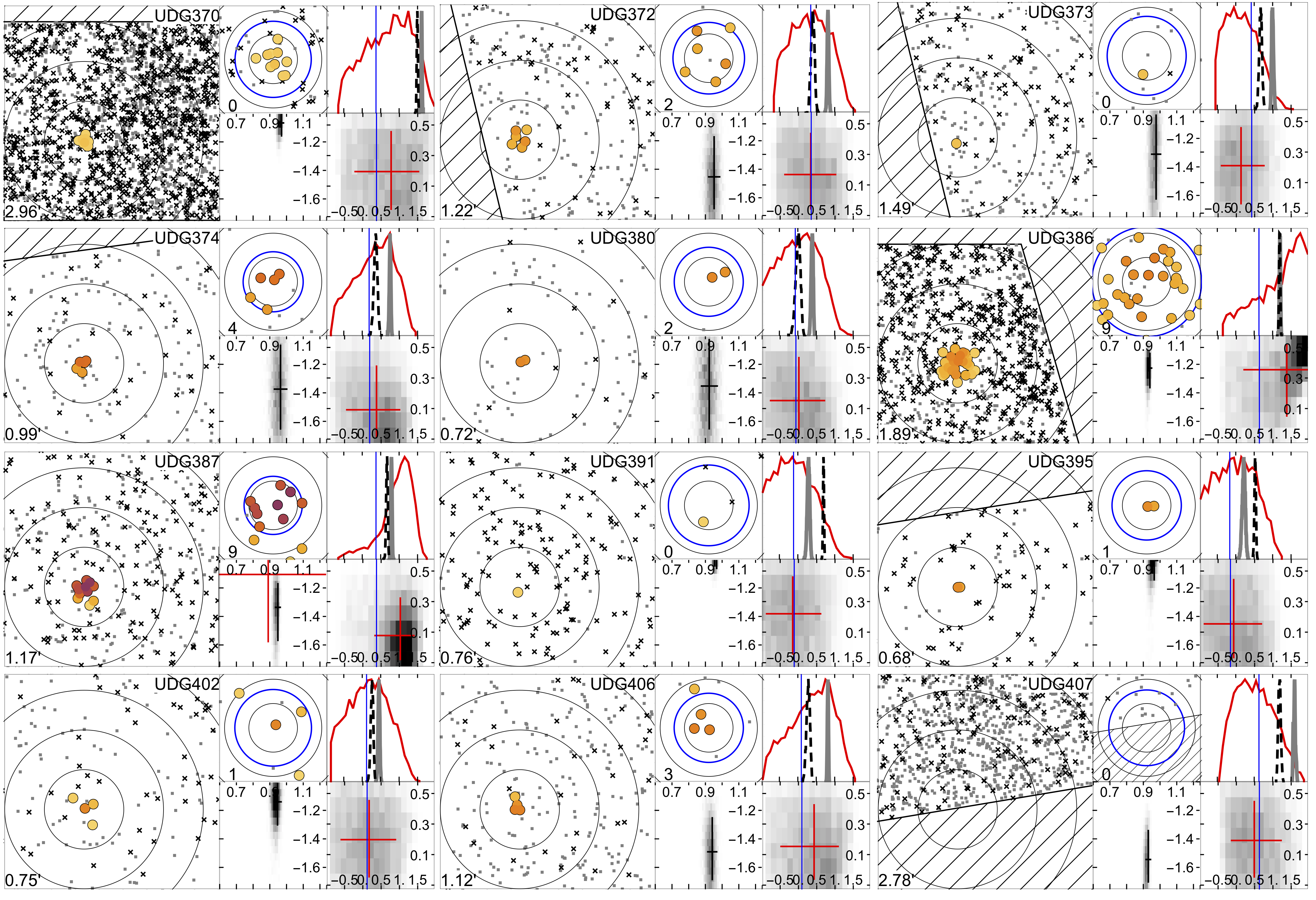}
\end{figure}
\end{landscape}
\begin{landscape}
\begin{figure}
\centering
\includegraphics[width=\columnwidth]{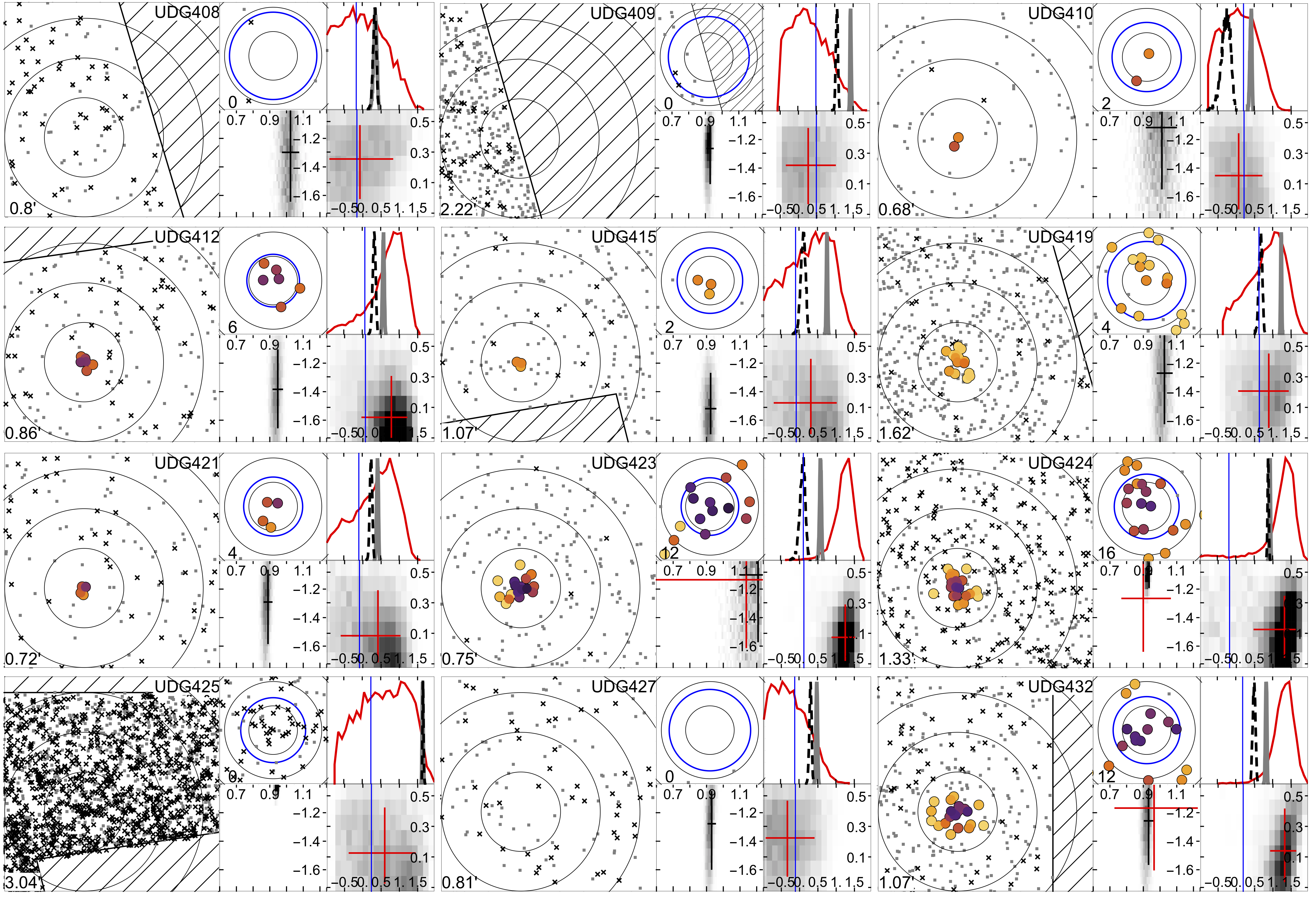}
\end{figure}
\end{landscape}
\begin{landscape}
\begin{figure}
\centering
\includegraphics[width=\columnwidth]{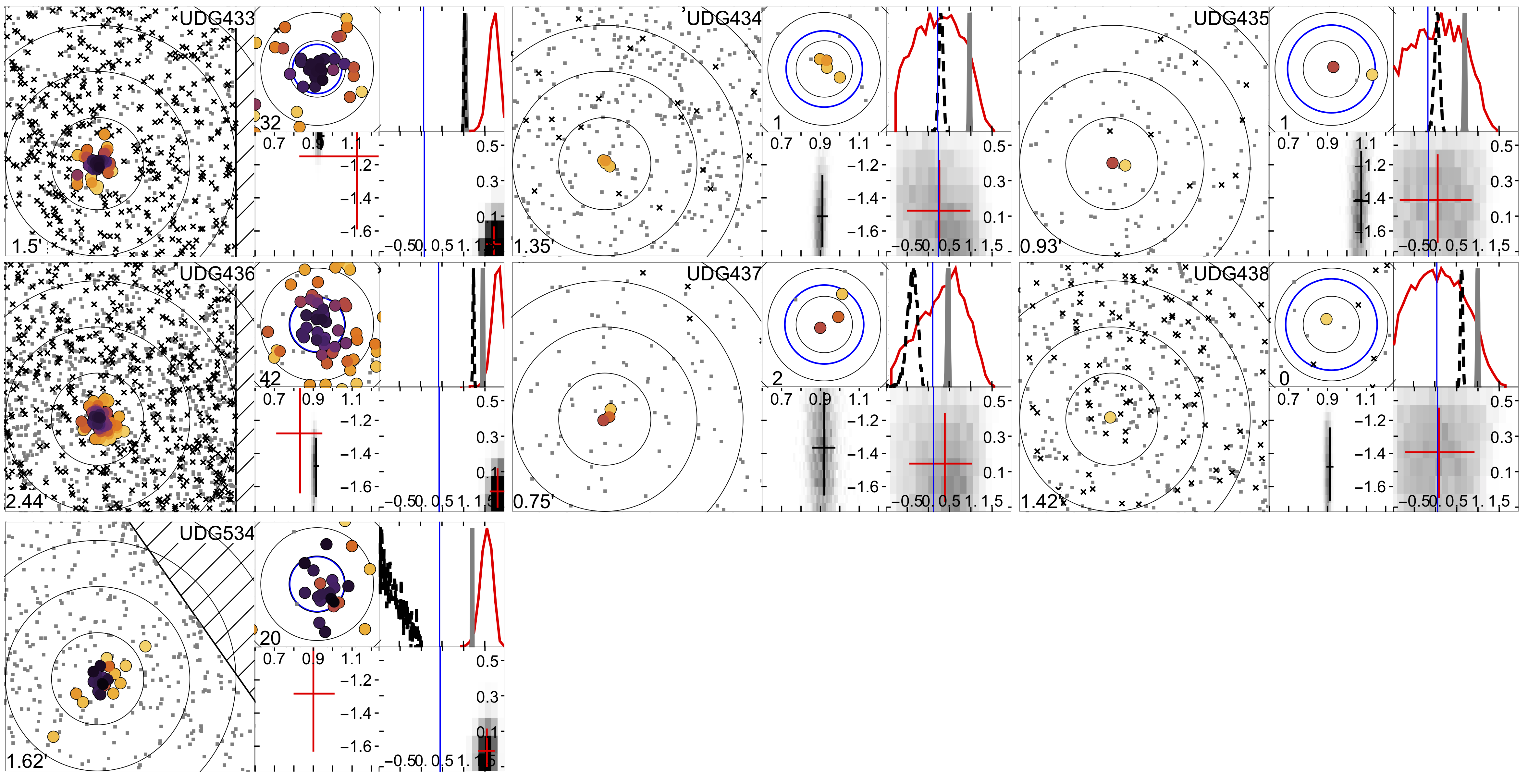}
\caption{Results of the analyses on individual systems. The ID number of each UDG is in the upper-right of panels {\it a}. 
Panels {\it a} and {\it b} zoom on Fig.~1, around each UDG. Black concentric circles display $\{5, 10, 15, 20\}\times R_S$ 
in panels {\it a} and $\{1,2\}\times R_S$ in panels {\it b}, where $R_S$ is the stellar half-light radius of the UDG. 
Panels {\it a} have a size $l \times l$, where $l$ is indicated in the lower-left of the panel itself, in arcmin. 
Where visible, the oblique black masking shows areas that lie outside the CCTp coverage and areas that are excluded 
due to bright galaxies. Each source in H10 is displayed according to its probability of membership in the UDG GCS.
Grey squares and black crosses are contaminants, respectively extended background galaxies and ICGCs. 
Filled circles are candidate GCs, $p_{\rm memb}>0.1$, colour-coded by $p_{\rm memb}$ itself, as shown in the legend. 
The blue circle in panels {\it b} shows the inferred value for the half-count radius of the UDG GC system.
Panels {\it c} and {\it d} show our inference on the UDG GC abundance, $N_{GC}$, after correcting by spatial and magnitude 
incompleteness, in red. The integer in the lower-left of panels {\it b} shows the number of GCCs with $p_{\rm memb}>0.25$.
 {Grey and black probability distribution functions in panels {\it c} show the GC abundance that would be inferred as a result 
of the contaminant population of resolved galaxies and ICGCs, respectively. }
The vertical blue line in the same panels shows the mean GC abundance of normal
dwarf galaxies with the same stellar mass, as from eqn.~(10).
Panels {\it e} display inferences on the colour properties of the ICGCs (grey shading and black cross) and, 
where at least 5 GCCs with {$p_{\rm memb}>0.5$} are identified, for the member GCs (red cross).
In both panels {\it d} and {\it e}, crosses extend between the 10\% and 90\% quantiles of the relevant 
posterior distributions.}
\label{mosaic}
\end{figure}
\end{landscape}

\section{Results}


Results for all our 54 UDGs are presented in Fig.~\ref{mosaic} and Table~1, which collects 10\%, 50\% and 90\% 
quantiles of the posterior distributions.
Results for our hyper parameters are collected in Table~2.
Fig.~\ref{mosaic} shows a mosaic of 54 panels, dedicated to each of our 54 UDGs.
Each panel is composed of insets {\it a} to {\it e}, as indicated by the legend (first panel). 

Insets {\it a} display zooms of Fig.~\ref{ACS} which include the central regions of each UDG
but are placed off-centre in order to allow a better impression of the statistics of the contaminants.
Coordinates in these panels are scaled to the UDG's effective radius $R_S$, and the concentric 
black circles display 5, 10, 15 and 20$\times R_S$, for scale. Panels {\it b} 
focus on the innermost regions of the UDG, where members are concentrated when present.
Black concentric circles in panels {\it b} display 1 and 2$\times R_S$, while the blue circle shows our inference
for the half-number radius of the UDG's GCS. In both panels {\it a} and {\it b},
each catalogue source is displayed with a symbol that codes its probability of membership 
in the UDG GCS, $p_{\rm memb}$, as obtained from our hierarchical Bayesian analysis. Sources that are contaminants with  
high probability ($p_{\rm memb}\leq10\%$) are displayed as a grey square or black cross based on whether
they are more likely background galaxies or ICGCs. The remainder are shown as filled circles, 
colour-coded by their individual probability of membership, as shown by the legend. 
As mentioned earlier and clear in panels {\it a} background counts are important in all studied systems. 
In the majority of cases, the number of background galaxies within $2\times R_S$ is higher 
than the number of GCCs identified as high probability members. Only in a fraction of cases the distribution of sources
shows obvious over-densities close to the centre of the UDG. The integer in the lower-left of insets {\it b} displays the 
number of detected GCCs that have $p_{\rm memb}>0.25$. 

Insets {\it c} display in red the marginalized posterior distribution for the richness of the GCS, $N_{GC}$,
after correcting for both spatial and magnitude incompleteness, as in~(\ref{incompl}). For convenience, the 
{probability distribution functions (pdfs)} in these
insets are normalised so that they peak at 1.
The vertical blue line shows the expected GC abundance based on the UDG stellar mass, $N_{GC,0}$, according to the 
empirical relation 
\begin{equation}
\log N_{GC,0} =  0.58 \log (M^*/M_\odot) - 4.09 \ ,
\label{normal}
\end{equation}
which provides the best linear fit in log-log space to the sample of \citet{IG08, IG09,IG10}.
This collects $>50$ nearby dwarf galaxies with HST imaging (see Fig.~\ref{stmass}). 
To enable use of the relation above, stellar masses for all our UDGs are obtained assuming that $B-R\sim1$ for the UDGs \citep{Koda15}, 
and averaging between the $M/L$ relations of \citet{Zi09} (see their Table~B1) and \cite{Be03} (see their Table~7).
We return on the comparison between our inferred GC abundances and those 
that would be predicted by the empirical relation~(\ref{normal}) in Section~4.2.

The grey-scale in insets {\it d} shows the joint probability distribution for $N_{GC}$ and 
$\log R_{\rm h}/R_S$, with a red cross extending over the 10\% and 90\% quantiles of either 
posterior distributions. In most cases, the half-number radius $R_{\rm h}$ is undetermined. These correspond
to the majority of cases in which there is no evidence for a central overdensity, and no high-probability 
candidate member is identified. In these instances, our inference on $N_{GC}$ is essentially an upper limit,
while the inference on $R_h$ is mostly determined by the prior volume (see Appendix A). 
In turn, for a fraction of the UDGs, we clearly detect a system of member GCs, corresponding to well-defined
bounds on both $N_{GC}$ and $\log R_{\rm h}/R_S$. It is worth noting that  
in the cases with apparently well detected GCSs, the preferred value of $R_h$ is often smaller than $R_S$ {(see
Table 1). This is even more striking considering that our prior for the ratio $R_h/R_S$ does not allow
values $<0.75$ (or $\log R_h/R_S<-0.125$).}

For comparison, panels {\it c} also show marginalized posterior distribution for 
background galaxies (in grey) and ICGCs (black dashed). These report inferences for the quantities
\begin{equation}
N_{{ICGC}}=N f_{ICGC}\times  {{\int \Sigma(r,R_{\rm h}/R_S=2)}\over{\int S_{{\rm sp}}({\bf r})\Sigma(r,R_{\rm h}/R_S=2)}} \mathcal{I}_{814}\ ,
\end{equation}
\begin{equation}
N_{{\rm gal}}=N (1-f_{GC}-f_{ICGC})\times  {{\int \Sigma(r,R_{\rm h}/R_S=2)}\over{\int S_{{\rm sp}}({\bf r})\Sigma(r,R_{\rm h}/R_S=2)}}\ ,
\end{equation}
where $\mathcal{I}_{814}$ is the completeness correction
\begin{equation}
\mathcal{I}_{814}={{\int g_{GC}(F814W)}\over{\int S_{814}g_{GC}(F814W)}} \ .
\end{equation}
$N_{{ICGC}}$ and $N_{{\rm gal}}$ are therefore the number of members that would be inferred
for that UDG as a consequence of the background counts in that field, statistically and assuming $R_{\rm h}/R_S=2$.
In an analysis in which one would simply count sources in a given aperture around the UDG centre, 
the grey and black pdfs show inference for the contamination that should be subtracted. 
By comparing with the posterior for $N_{GC}$, it is clear that background counts are largely dominant
in most systems, making this simple method to estimate GC abundances prone to significant uncertainties.
Only those systems in which $N_{{ICGC}}+N_{{\rm gal}}$ is significantly smaller than the inferred $N_{GC}$ 
do indeed display a significant overdensity associated with the UDG GCS.

Finally, panels {\it e} address the colour properties of ICGCs and member GCs. {
It is worth noticing that, throughout the paper, results on the colour distribution of our populations of 
point sources -- the population of member GCs and the ICGC population -- take into account the filter dependence 
of the aperture correction, which we calculate using \citet{Sirianni05}, $\delta\textsc{c}=0.09$~mag. {
By comparing with the colour in a larger aperture, we estimate that resolved galaxies require a correction to their colour 
$\delta\textsc{c}_{\rm gal}\lesssim0.01$~mag, which we therefore decide not to apply in Table~2.} For ICGCs and member GCs, 
inferences are displayed only when the number of members with probability $p>0.5$ 
in either populations is larger than 5. The grey-scale shows the 
joint posterior distribution for the mean and dispersion of the colour of the ICGCs in each field, with the black cross
ranging between the 10\%-90\% quantiles. Both mean colour and spread of the ICGCs vary from field to field, 
although they remain typical for GCs \citep[see e.g.,][]{Pe08,Pe16,MB16b}. Note that, the only fields in which we 
cannot constrain the colour properties of the ICGCs are UDGs 236, 238 and 534. As shown by Fig.~1, these galaxies 
lie in the CCTp fields that are further away from the centre of Coma, and therefore the surface density of ICGCs is 
expected to be lower there \citep{Pe11}, in line with our findings. {The fact we do not identify 
high probability members of the ICGC population in these fields implies that the number of foreground stars and 
background galaxies that `contaminate' this population is negligible. The large majority of the high probability members of 
the ICGC population are indeed bona fide ICGCs.}
Unfortunately, we find it very difficult to constrain the colour properties of the different GCSs. 
The red crosses in panels {\it e}
show 10\%-90\% quantiles, but, even for those cases in which member GCs are securely identified, inferences remain 
rather inconclusive. We return on the colour properties of the member GCs in Section~\ref{GCcolour}.

With reference to possible galactic nuclei, we consider the minimum distance of any high-probability members into the UDG GCS.
When candidates with $p_{\rm memb}>0.5$ exist within 150~pc from the UDG centre, Table~1 records the minimum galactocentric 
distance $R_{\rm min}$. There are 4 systems with high-probability members with $R_{\rm min}\leq100$~pc, and an additional
5 with $100<R_{\rm min}/{\rm pc}<150$.

 \begin{table*}
  \caption{Summary of the statistical constraints from our analyses. 
  Column~1 lists the UDGs in our sample by their ID number, as in the catalogue by \citet{Ya16}; 
  columns~{2} and~{3} list stellar scale radii and stellar masses; 
  column~4 collects the GC abundances expected based on the stellar mass of the system, according to eqn.~(\ref{normal}); 
  column~5 lists inferences on the GC abundance; 
  {column~6 records the ratio $R_h/R_S$;}
  columns~7 and~8 refer to the properties of the colour distribution of the ICGCs in that field;
  column~9 uses inference on the GC abundance to produce upper limits (90\% quantiles) for the virial mass of each UDG, $M_{vir,90}$;
  when GC candidates with probability of membership into the UDG GCS $p_{\rm memb}>0.5$ exist within 150~pc from the UDG centre, 
  column~10 lists the minimum galactocentric distance of these possible nuclei.
  Columns featuring a trio of entries collect the $\{10\%,50\%,90\%\}$ of the relevant marginalised posterior distribution.   }
  \label{res-table}
  \begin{tabular}{cccc cccc c c}
  
 ID &   $R_S$     &  $\log M_*$& $N_{GC,0}$ & $N_{GC}$    & $\log R_h/R_S$  & $\langle$colour$\rangle_{ICGC}$ & $\sigma($colour$_{ICGC})$  & $M_{\rm vir,90}$  & $R_{\rm min}$\\ 
       & $\left[ {\rm kpc}\right]$ &            $\left[M_\odot \right]$ &          &        &      &$\left[{\rm mag} \right]$            &             $\left[{\rm mag} \right]$ &  $\left[10^{11}M_{\odot}\right]$ &   pc  \\
\hline
85&0.94&7.18&1.2&$\{0.2, 0.7, 3.4\}$&$\{-0.06, 0.18, 0.45\}$&$\{0.88, 0.90, 0.93\}$&$\{0.02, 0.03, 0.05\}$&0.17& --\\
89&1.10&7.50&1.9&$\{0.5, 3.9, 22.1\}$&$\{0.01, 0.29, 0.49\}$&$\{0.96, 0.98, 1.01\}$&$\{0.03, 0.06, 0.10\}$&1.3& --\\
91&1.58&6.78&0.7&$\{0.1, 0.5, 3.3\}$&$\{-0.03, 0.21, 0.46\}$&$\{0.88, 0.91, 0.94\}$&$\{0.02, 0.04, 0.07\}$&0.16& --\\
99&2.09&7.22&1.3&$\{0.5, 14.8, 70.5\}$&$\{0.03, 0.32, 0.50\}$&$\{0.90, 0.91, 0.92\}$&$\{0.03, 0.05, 0.07\}$&5.0& --\\
102&0.89&7.02&1.0&$\{5.0, 11.5, 21.0\}$&$\{-0.11, -0.04, 0.11\}$&$\{0.89, 0.92, 0.96\}$&$\{0.02, 0.05, 0.09\}$&1.3& 122\\
104&1.05&7.30&1.5&$\{9.8, 18.7, 31.0\}$&$\{-0.11, -0.03, 0.10\}$&$\{0.87, 0.91, 0.94\}$&$\{0.02, 0.04, 0.07\}$&2.0& 38\\
105&1.22&6.58&0.6&$\{0.1, 0.5, 3.0\}$&$\{-0.05, 0.20, 0.44\}$&$\{0.89, 0.91, 0.92\}$&$\{0.02, 0.03, 0.05\}$&0.14& --\\
107&1.99&6.78&0.7&$\{0.1, 0.8, 5.4\}$&$\{-0.04, 0.20, 0.45\}$&$\{1.00, 1.02, 1.04\}$&$\{0.02, 0.04, 0.07\}$&0.27& --\\
108&0.81&6.58&0.6&$\{0.1, 0.4, 2.2\}$&$\{-0.04, 0.22, 0.47\}$&$\{0.81, 0.87, 0.92\}$&$\{0.02, 0.05, 0.10\}$&0.10& --\\
112&1.53&7.74&2.6&$\{0.4, 1.3, 6.2\}$&$\{-0.01, 0.25, 0.47\}$&$\{0.94, 0.96, 0.98\}$&$\{0.04, 0.07, 0.09\}$&0.32& --\\
113&0.84&6.38&0.4&$\{0.1, 0.6, 3.8\}$&$\{-0.05, 0.19, 0.45\}$&$\{0.91, 0.94, 0.97\}$&$\{0.02, 0.03, 0.06\}$&0.19& --\\
114&1.52&7.74&2.6&$\{0.4, 1.7, 6.9\}$&$\{-0.05, 0.19, 0.44\}$&$\{0.89, 0.91, 0.93\}$&$\{0.02, 0.03, 0.07\}$&0.36& --\\
115&1.16&6.58&0.6&$\{0.3, 5.0, 19.7\}$&$\{-0.06, 0.15, 0.42\}$&$\{0.89, 0.92, 0.95\}$&$\{0.02, 0.04, 0.07\}$&1.2& --\\
118&0.88&6.82&0.8&$\{0.1, 0.6, 3.2\}$&$\{-0.03, 0.23, 0.46\}$&$\{0.90, 0.92, 0.95\}$&$\{0.03, 0.05, 0.09\}$&0.15& --\\
121&1.67&7.82&2.9&$\{0.5, 1.5, 6.6\}$&$\{-0.03, 0.22, 0.46\}$&$\{0.93, 0.96, 0.99\}$&$\{0.04, 0.08, 0.12\}$&0.35& --\\
122&1.47&7.58&2.1&$\{1.3, 10.1, 26.5\}$&$\{-0.08, 0.10, 0.38\}$&$\{0.94, 0.97, 1.00\}$&$\{0.02, 0.05, 0.10\}$&1.7& --\\
236&1.25&6.86&0.8&$\{1.9, 8.6, 20.6\}$&$\{-0.10, 0.02, 0.23\}$&-- & --&1.2& 121\\
238&1.06&7.34&1.5&$\{0.3, 1.2, 6.0\}$&$\{-0.03, 0.22, 0.46\}$&-- & --&0.31& --\\
325&1.01&7.02&1.0&$\{13.9, 25.0, 40.8\}$&$\{-0.09, 0.05, 0.27\}$&$\{0.83, 0.87, 0.91\}$&$\{0.02, 0.04, 0.08\}$&2.7& 43\\
331&1.47&7.58&2.1&$\{0.3, 1.1, 4.8\}$&$\{-0.04, 0.21, 0.46\}$&$\{0.89, 0.91, 0.92\}$&$\{0.02, 0.03, 0.06\}$&0.24& --\\
358&1.75&8.10&4.3&$\{1.0, 5.4, 18.4\}$&$\{-0.06, 0.13, 0.39\}$&$\{0.87, 0.89, 0.91\}$&$\{0.02, 0.03, 0.05\}$&1.1& --\\
366&1.57&7.74&2.6&$\{0.4, 1.3, 5.0\}$&$\{-0.04, 0.19, 0.44\}$&$\{0.94, 0.96, 0.98\}$&$\{0.04, 0.08, 0.10\}$&0.25& --\\
367&1.48&7.58&2.1&$\{0.5, 4.3, 20.5\}$&$\{-0.04, 0.20, 0.43\}$&$\{0.94, 0.96, 0.97\}$&$\{0.03, 0.06, 0.09\}$&1.2& --\\
370&3.03&7.66&2.4&$\{0.6, 5.9, 33.3\}$&$\{-0.05, 0.20, 0.46\}$&$\{0.95, 0.96, 0.97\}$&$\{0.07, 0.09, 0.10\}$&2.1& --\\
372&1.25&7.62&2.2&$\{0.4, 2.2, 10.3\}$&$\{-0.06, 0.17, 0.44\}$&$\{0.91, 0.94, 0.98\}$&$\{0.02, 0.04, 0.07\}$&0.57& --\\
373&1.52&7.74&2.6&$\{0.4, 1.3, 5.7\}$&$\{-0.04, 0.21, 0.46\}$&$\{0.94, 0.97, 0.99\}$&$\{0.02, 0.05, 0.09\}$&0.29& --\\
374&1.01&7.34&1.5&$\{0.4, 2.4, 10.1\}$&$\{-0.08, 0.09, 0.38\}$&$\{0.92, 0.95, 0.99\}$&$\{0.02, 0.04, 0.09\}$&0.56& --\\
380&0.74&6.90&0.9&$\{0.2, 1.0, 5.1\}$&$\{-0.06, 0.15, 0.43\}$&$\{0.87, 0.91, 0.96\}$&$\{0.02, 0.04, 0.09\}$&0.26& --\\
386&1.93&8.10&4.3&$\{1.6, 24.0, 86.6\}$&$\{0.05, 0.35, 0.52\}$&$\{0.92, 0.93, 0.94\}$&$\{0.04, 0.06, 0.08\}$&6.2& --\\
387&1.20&7.66&2.4&$\{2.2, 10.6, 24.8\}$&$\{-0.08, 0.08, 0.32\}$&$\{0.92, 0.94, 0.95\}$&$\{0.03, 0.05, 0.07\}$&1.5& --\\
391&0.78&6.82&0.8&$\{0.1, 0.7, 4.0\}$&$\{-0.04, 0.22, 0.46\}$&$\{0.94, 0.96, 0.97\}$&$\{0.09, 0.10, 0.12\}$&0.20& --\\
395&0.69&6.74&0.7&$\{0.1, 0.8, 4.8\}$&$\{-0.07, 0.15, 0.44\}$&$\{0.94, 0.96, 0.98\}$&$\{0.07, 0.10, 0.12\}$&0.24& --\\
402&0.77&7.22&1.3&$\{0.3, 1.5, 7.8\}$&$\{-0.04, 0.20, 0.45\}$&$\{0.91, 0.94, 0.96\}$&$\{0.05, 0.07, 0.10\}$&0.42& --\\
406&1.15&7.18&1.2&$\{0.3, 2.7, 12.1\}$&$\{-0.06, 0.15, 0.43\}$&$\{0.90, 0.93, 0.96\}$&$\{0.02, 0.03, 0.06\}$&0.69& --\\
407&2.85&8.10&4.3&$\{0.7, 3.1, 16.7\}$&$\{-0.05, 0.19, 0.45\}$&$\{0.91, 0.92, 0.93\}$&$\{0.02, 0.03, 0.05\}$&0.98& --\\
408&0.82&6.74&0.7&$\{0.1, 0.8, 6.4\}$&$\{-0.00, 0.25, 0.47\}$&$\{0.97, 1.01, 1.06\}$&$\{0.02, 0.05, 0.11\}$&0.34& --\\
409&2.27&7.86&3.1&$\{0.5, 1.8, 10.0\}$&$\{-0.03, 0.22, 0.46\}$&$\{0.90, 0.92, 0.94\}$&$\{0.03, 0.05, 0.08\}$&0.55& --\\
410&0.69&7.38&1.6&$\{0.3, 1.2, 4.9\}$&$\{-0.06, 0.15, 0.43\}$&$\{0.91, 1.00, 1.09\}$&$\{0.03, 0.08, 0.17\}$&0.25& --\\
412&0.88&7.14&1.2&$\{1.0, 6.0, 15.7\}$&$\{-0.09, 0.03, 0.30\}$&$\{0.91, 0.94, 0.96\}$&$\{0.02, 0.04, 0.08\}$&0.92& --\\
415&1.10&6.86&0.8&$\{0.2, 2.0, 9.9\}$&$\{-0.07, 0.13, 0.41\}$&$\{0.89, 0.91, 0.94\}$&$\{0.02, 0.03, 0.05\}$&0.54& --\\
419&1.66&8.10&4.3&$\{1.2, 7.7, 25.8\}$&$\{-0.03, 0.21, 0.45\}$&$\{0.97, 1.02, 1.06\}$&$\{0.02, 0.05, 0.11\}$&1.6& --\\
421&0.74&6.86&0.8&$\{0.3, 2.6, 10.2\}$&$\{-0.08, 0.08, 0.38\}$&$\{0.86, 0.88, 0.90\}$&$\{0.03, 0.05, 0.09\}$&0.56& --\\
423&0.77&7.22&1.3&$\{7.9, 17.7, 31.5\}$&$\{-0.08, 0.07, 0.28\}$&$\{1.10, 1.20, 1.30\}$&$\{0.03, 0.08, 0.20\}$&2.0& 131\\
424&1.36&6.70&0.7&$\{3.1, 20.7, 40.0\}$&$\{-0.06, 0.13, 0.34\}$&$\{0.91, 0.92, 0.94\}$&$\{0.07, 0.08, 0.10\}$&2.6& --\\
425&3.11&7.42&1.7&$\{0.4, 4.0, 26.7\}$&$\{-0.07, 0.13, 0.42\}$&$\{0.93, 0.94, 0.95\}$&$\{0.09, 0.10, 0.11\}$&1.7& --\\
427&0.83&6.82&0.8&$\{0.1, 0.5, 2.4\}$&$\{-0.02, 0.22, 0.46\}$&$\{0.90, 0.92, 0.94\}$&$\{0.03, 0.05, 0.09\}$&0.11& --\\
432&1.10&7.34&1.5&$\{8.8, 21.0, 41.4\}$&$\{-0.06, 0.14, 0.41\}$&$\{0.90, 0.92, 0.95\}$&$\{0.03, 0.06, 0.09\}$&2.7& 52\\
433&1.54&7.14&1.2&$\{34.8, 51.2, 71.1\}$&$\{-0.11, -0.06, 0.04\}$&$\{0.92, 0.93, 0.94\}$&$\{0.08, 0.09, 0.11\}$&5.0& 148\\
434&1.38&7.42&1.7&$\{0.3, 1.8, 9.2\}$&$\{-0.07, 0.13, 0.41\}$&$\{0.88, 0.90, 0.92\}$&$\{0.02, 0.03, 0.05\}$&0.50& --\\
435&0.95&6.74&0.7&$\{0.2, 1.1, 6.7\}$&$\{-0.04, 0.19, 0.44\}$&$\{1.03, 1.06, 1.10\}$&$\{0.02, 0.04, 0.08\}$&0.35& 94\\
436&2.50&7.74&2.6&$\{43.0, 62.6, 87.6\}$&$\{-0.10, -0.01, 0.11\}$&$\{0.90, 0.90, 0.91\}$&$\{0.02, 0.03, 0.05\}$&6.3& --\\
437&0.77&7.22&1.3&$\{0.4, 2.4, 10.0\}$&$\{-0.07, 0.15, 0.43\}$&$\{0.85, 0.91, 0.96\}$&$\{0.02, 0.04, 0.09\}$&0.55& 141\\
438&1.45&7.10&1.1&$\{0.2, 1.2, 7.9\}$&$\{-0.04, 0.21, 0.46\}$&$\{0.89, 0.90, 0.92\}$&$\{0.02, 0.03, 0.06\}$&0.42& --\\
534&1.66&7.78&2.8&$\{23.0, 35.0, 51.2\}$&$\{-0.10, -0.01, 0.11\}$&-- & --&3.5& --\\
    \hline
  \end{tabular}
 \end{table*}

\begin{table}
  \caption{Inference on the model hyper-parameters.}
  \label{res-table2}
  \centering
  \begin{tabular}{cc}
  \hline
 $(\langle C_{1.2-4}\rangle, \sigma(C_{1.2-4}))_{GC}$ & $(1.07, 0.042)$~mag\\
 $(\langle C_{1.2-4}\rangle, \sigma(C_{1.2-4}))_{\rm gal}$ & $(1.55, 0.51)$~mag\\
  \hline
 $(\langle \textsc{c}\rangle, \sigma(\textsc{c}))_{GC}$ & $(0.91, 0.06)$~mag\\
 $(\langle \textsc{c}\rangle, \sigma(\textsc{c}))_{ICGC}$ & $(0.93, 0.07)$~mag\\
  $(\langle \textsc{c}\rangle, \sigma(\textsc{c}))_{\rm gal}$ & $({0.92}, 0.57)$~mag\\
  \hline
  \end{tabular}
   \end{table}

\subsection{Upper limits for individual virial masses}

We follow \citet{Ha15}, and use GC abundance as a proxy for virial mass.
In particular, we interpret value of the 90\% quantile of the posterior distribution for $N_{GC}$, 
$N_{GC,90}$ as an upper limit for the virial mass of each UDG, $M_{vir,90}$. In order to do so, 
we adopt the calibration presented by \citet{Ha17}, assuming its validity extends to the regime of UDGs:
\begin{equation}
\begin{cases}
{M_{\rm vir}} = N_{GC}/\eta_N  \\
\log \eta_N = -8.56-0.11\log M_{vir}/M_\odot 
\end{cases}\ .
\end{equation}
Results are listed in the second to last column of Table~1. In our sample of 54 systems, 33 galaxies have a virial mass
$M_{vir}\leq10^{11}M_\odot$ at 90\% probability. The remaining 21 are approximately equally split between
systems with $M_{vir,90}\leq 4\times 10^{11}M_\odot$ and systems with higher 90\% upper limits.
Note, however, that among these 21, 18 systems have GC abundances that are essentially undetermined because of the
high background counts, with values of $M_{vir,10}$ below $10^{11}~M_\odot$ also allowed. For only three galaxies we infer
halo masses in excess of  $10^{11}~M_\odot$ at 90\% probability.
Interestingly, among the 21 galaxies with $M_{vir,90}> 10^{11}M_\odot$, 12 have a half-light radius $R_S<1.5$~kpc,
implying that 9 of the 18 UDGs with $R_S>1.5$~kpc have $M_{vir}\leq10^{11}M_\odot$ at 90\% probability.

\begin{figure}
\centering
\includegraphics[width=\columnwidth]{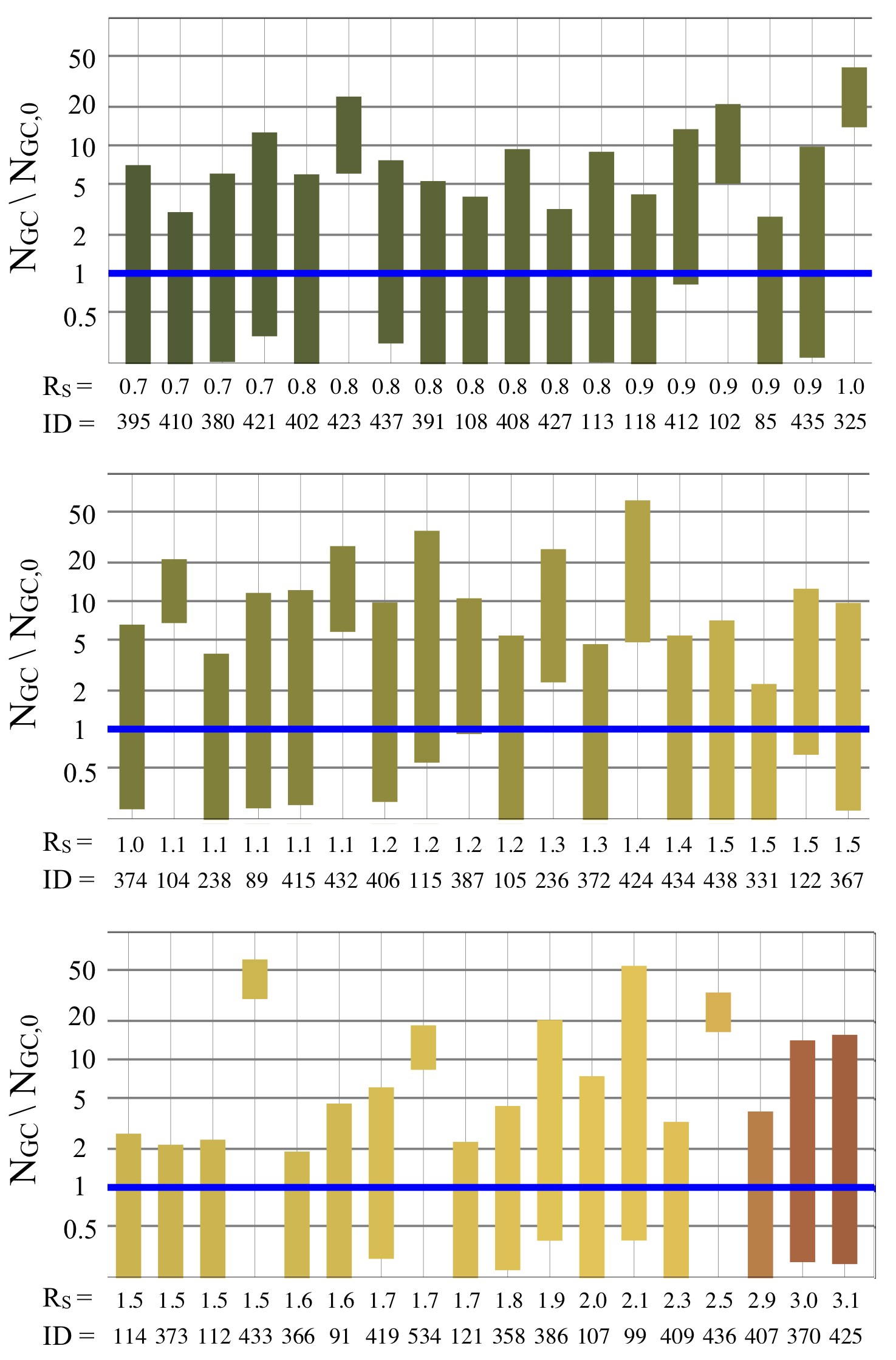}
\caption{The richness of the UDG GC systems in terms of the mean richness of `normal' dwarf galaxies with the same stellar mass, $N_{GC}/N_{GC,0}$.
The 54 UDGs in our sample are ordered by size, and colour-coded accordingly. Bars extend between the 10\% and 90\% quantiles.}
\label{relativeN}
\end{figure}

\subsection{Comparison with `normal' dwarfs}

Figure~\ref{relativeN} illustrates how the GC abundances we measure compare with 
the values expected based on the UDG stellar masses. 
Coloured bars extend between the 10\% and 90\% quantiles of 
the quantity $N_{GC}/N_{GC,0}$, where $N_{GC,0}$ is according to the relation~(\ref{normal}).
UDGs are ordered (and colour-coded) by the stellar half-light radius $R_S$. 
Our results show that most galaxies in our sample have `normal' GCSs for their stellar mass:
the majority of the low surface brightness galaxies we can study are consistent the haloes of `normal' dwarf
galaxies. According to Fig.~\ref{relativeN}, 10 systems display GCSs that are `overabundant' with respect to the mean
relation at more than 90\% probability; 9 galaxies have $N_{GC}/N_{GC,0}>3$ at 90\% probability. 
It is unclear, however, whether this is particularly surprising given the substantial scatter about the relation
(see Figure 6 below). We return on this aspect in Section~5.
It is interesting to notice that while three of the 9 galaxies that satisfy $N_{GC,10}/N_{GC,0}>3$ 
have $R_S>1.5$~kpc, the remaining 6 have smaller sizes, with 5 having  $R_S\lesssim1.1~{\rm kpc}$.

\begin{figure}
\centering
\includegraphics[width=.9\columnwidth]{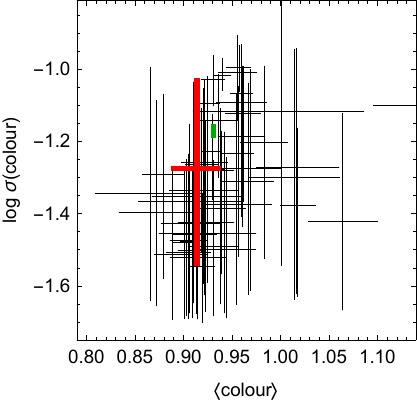}
\caption{The colour properties inferred for the ensemble of UDG GCs (red cross, extending between 10\% and 90\% of
the probability distribution for either mean colour and spread). These are compared with the properties of the ICGCs 
in each individual UDG field (thin black crosses), as well as with the properties of the ensemble of ICGCs in all
studied fields ({green rectangle}, extending between 10\% to 90\% quantiles
of the inferred probability distributions). Inferences are also reported in Table~2.}
\label{colourcomp}
\end{figure}

\subsection{The colour of member GCs}\label{GCcolour}

As discussed in Section 4, the quality of the available data and the limited
number of members do not, in most cases, allow for useful constraints on the 
colour properties of the individual UDG GCSs. We therefore perform a parallel analysis in
which we assume that all of the UDG GCSs share the same mean colour $\langle \textsc{c}\rangle_{GC}$ 
and dispersion $\sigma(\textsc{c})_{GC}$, and elevate both to hyper parameters.
This allows for improved constraints by using all of the detected high probability members
over the 54 UDGs. We record our inference in Table~2 and display 10\% to 90\% quantiles of the 
joint probability distribution as a red cross in Figure~\ref{colourcomp}. 
The same Figure collates results obtained for the same parameters for 
the ICGCs in each of the 54 studied fields, shown as thin black crosses. 
Most of these are perfectly compatible with what we find for the UDG GCSs.
For a better comparison, we also elevate the properties of the 
ICGCs to hyper-parameters, to measure the mean properties of the 
ICGCs in all of Coma. Given the large number of ICGCs identified in our analysis, 
their properties are very well determined. We display results in Fig.~\ref{colourcomp}
{as a green rectangle}, extending between 10\% to 90\% quantiles
of the inferred probability distributions (mean values are appended in Table~2).
Results obtained for the ensemble of the member GCs and for the ensemble of the ICGCs
are perfectly compatible, in both mean colour and spread. The available data do not allow to distinguish
the mean properties of the two populations. In other words, we cannot rule out that GC lost by 
disrupted UDGs make up for most of the ICGC population.

\subsection{Comparison with literature work}
Concurrently with this study, \citet{vD17} (hereafter vD17) published an independent analysis 
that includes 12 of the galaxies in our sample, based on the same CCTp imaging used to produce 
the H10 catalogue. The parallel study by vD17, however, is different in that they perform
their own source extraction, after explicitly fitting for and subtracting the 12 studied UDGs
from the images themselves (rather than by correcting the aperture photometry after the source 
extraction as we do in Section~2.1). In addition, the vD17 analysis and the present one differ
substantially in methodology, using two different techniques to assess GC abundances.
While we perform a statistical analysis of all sources in large fields around each UDG, 
\citet{vD17} adopts the approach of counting all sources within an aperture centred on the 
UDG to then subtract an estimate of the background contamination. 
This number count is then corrected for magnitude incompleteness (a factor of 2,
analogous to our factor $\mathcal{I}_{814}$) and by spatial incompleteness, 
assuming that all UDG GCSs have a half-number radius of $1.5\times R_S$.

Our results are in rough agreement with those presented by vD17,
with our uncertainties often wider than those recorded by that study.
The most discrepant results are for UDGs with ID 358, for which vD17 records
a lower 1-sigma limit of 31 while we find an upper limit of $N_{GC,90}=18.4$, 
and ID 121, with respectively 14 and 6.6. In turn, we find that the richness
of UDGs ID 122 and ID 386 have a high uncertainty because of the background counts,
but could potentially be significantly higher than estimated by vD17.

\section{Discussion and Conclusions}

\begin{figure*}
\centering
\includegraphics[width=\textwidth]{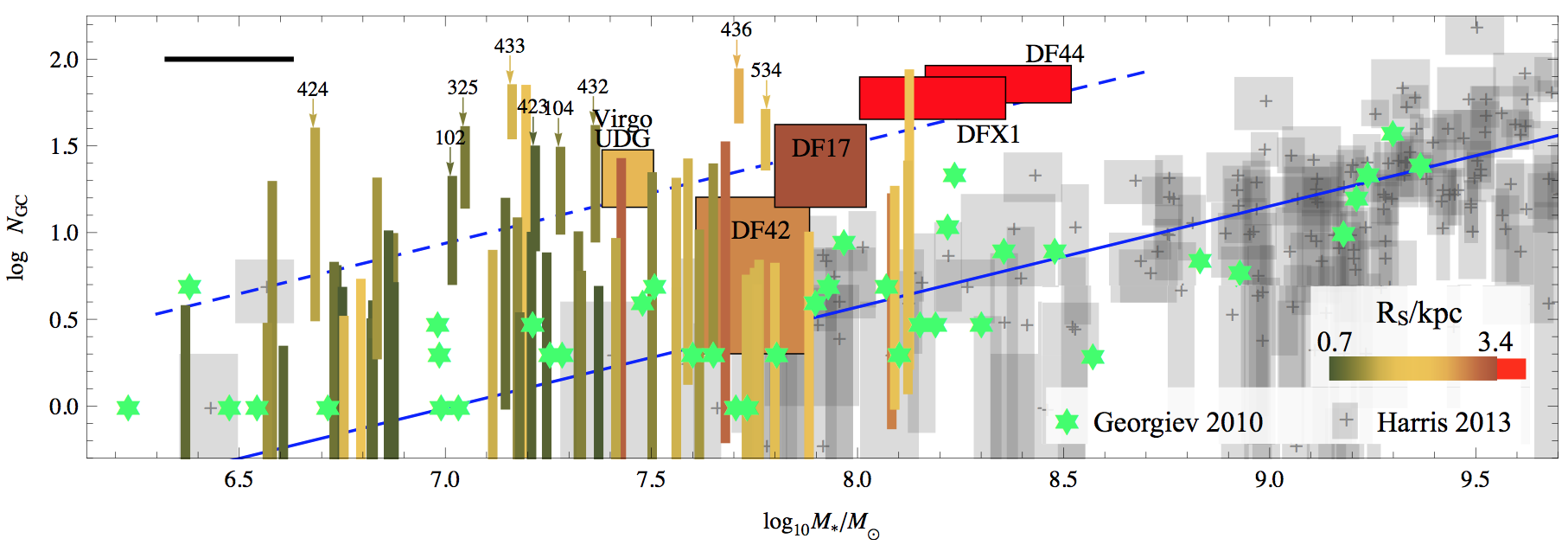}
\caption{The richness of the GCS of `normal' dwarf galaxies, from the H13 catalogue (grey rectangles) and from G10 (green stars).
The solid blue line is a fit to the properties of the dwarf galaxies collected by \citet{IG08,IG09,IG10}, recorded in eqn.~(\ref{normal}). 
Results for all our 54 UDGs (90\% confidence regions) are shown as coloured bars, colour-coded 
by the UDG stellar half-light radius. {The horizontal black bar in the upper-left
illustrates the size of the uncertainty on the UDG stellar masses. }
Literature measurements for UDG GC abundances are shown as full rectangles with 
a black edge, with the same colour-coding. The dashed blue line provides fit to the relation that characterises galaxies with 
`enhanced' GC richness with respect to their stellar mass, characterized by GC abundances approximately 9 times higher than normal galaxies. }
\label{stmass}
\end{figure*}

Our results show that the majority of UDGs are hosted by dwarf mass haloes, with 
$M_{vir}\lesssim10^{11} M_\odot$. For only 3 systems our inference on the richness of their 
GCS translates to a virial mass that is higher than $10^{11} M_\odot$ at 90\% confidence (UDGs 
with ID 436, 433 and 534, have $M_{vir,10}>10^{11} M_\odot $).
According to the bounds allowed by the currently available dataset, most UDG GCSs are in fact consistent
with expectations for normal dwarf galaxies with the same stellar mass.
A fraction of systems, however, display richer GCSs. This property does not 
seem to correlate with either stellar mass or galaxy size. We find
9 systems with GCS that are richer by more than 3 times (at 90\% probability) with respect to the
mean of the population of `normal' dwarfs with similar stellar mass. 
Of these, three galaxies have $R_S>1.5$~kpc, while the remaining six (UDGs with ID 102, 104, 325, 423, 424, 432)
are less extended, with half-light radii ranging between 0.8 and 1.4~kpc. 
These galaxies enlarge the sample of known dwarfs with especially abundant GCSs \citep[see e.g.,][]{JL04,Pe08},
adding nine low surface brightness systems. 

Our results suggest that the extended size and unusually high GC abundance do not 
necessarily accompany each other. The physical mechanism responsible for the uncommonly
high half-light radii of the UDGs has not yet been pinpointed unequivocally, but 
it appears that the `over-abundance' of some UDG GCS may in fact be unrelated.
In turn, this may call into question the threshold of $R_S>1.5$~kpc used so far to classify UDGs. 
Our sample of 54 low surface brightness galaxies includes extended UDGs ($R_S>1.5$~kpc) with normal
GCSs and relatively compact dwarfs with clearly elevated GC abundances. Certainly, all of our galaxies 
are well within the dwarf regime, we find no single system where GC abundances typical 
for MW like galaxies \citep[$N_{GC,MW}\approx 144$, ][]{Ha17} are allowed within 
the 90\% confidence region. 

In addition, it is worth noticing that, in those systems in which large values of $N_{GC}/N_{GC,0}$ are securely identified, 
our analysis suggests that $R_h/R_S\lesssim 2$ with high probability, with at least 4 cases in which $R_h/R_S\lesssim 1$.
Thus the GCSs in our objects have values of $R_h$ which resemble those
of normal galaxies of the same stellar mass rather than of the same size. Under the hypothesis that GCS properties 
are more closely related to dark matter halos than to central galaxies, this may suggest that halo extent (and hence mass) 
is more closely linked to stellar mass than to size in our objects. This may provide independent evidence that these objects 
do not have overmassive haloes compared to the expectation for their stellar mass, though deeper datasets, to better account for the high 
background counts, would be valuable.

\subsection{A separate cluster population of GC-rich galaxies?}

Fig.~\ref{stmass}, compares the inferred GC abundances of our 54 UDGs with those of nearby galaxies from the 
compilations of G10 (green stars) and \citet[][H13, grey rectangles]{Ha13}, as a function of stellar mass\footnote{
For the H13 catalogue, stellar masses were obtained using morphological type as a proxy for colour. For our UDGs, 
the uncertainty displayed in the top-left shows the difference between the masses inferred using the $M/L$ relations 
of \citet{Zi09} and \cite{Be03}.}.
The blue solid line represents the relation~(\ref{normal}), which, we recall, is a fit to the properties of the sample 
of nearby dwarf galaxies by \citet{IG08,IG09,IG10}. Note that this includes a minority of galaxies 
that are not displayed in Fig.~\ref{stmass}, as bearing no GCs. 
Vertical bars (colour-coded by stellar half-light radius) cover the range of GC abundances 
that are compatible with the CCTp data at 90\% confidence, while the horizontal black bar in the upper-left
illustrates the size of the uncertainty on the UDG stellar masses. Full rectangles display literature measurements
for other UDGs not included in our sample \citep{Pe16,MB16b,vD17}.

\citet{MB16b} and \citet{Pe16} have suggested that, though hosted by dwarf haloes with masses similar 
to that of the LMC, UDGs might still have rich GCSs for their stellar mass. This could 
be interpreted in two different ways: i) as a sign of `failure' in forming stars, i.e. as an especially low star formation efficiency or
as a consequence of premature gas removal or quenching of some form \citep{vD15,YB15,MB16b,Pe16}; ii)  
as the result of an especially high GC formation efficiency \citep[e.g.,][]{Pe08}. 

To address this, we perform a statistical analysis on the relation between stellar mass and GC abundance in our sample. 
We aim to establish whether the population of Coma low surface brightness galaxies, as probed by our 54 systems, can be
described by a single population of galaxies (with significant scatter in the relation between stellar mass and GC abundance), or 
whether the data suggest the existence of an additional population with especially rich GCSs. We do so by 
describing our results in the $(M^*, N_{GC})$ plane with a two component model, in which the first population 
is as from the relation~(\ref{normal}) while a second population has GCSs richer by a factor X for the same 
stellar mass (and the same slope).
Similarly to eqns.~(5-7), UDG $j$ has a probability of belonging into the population of normal galaxies of
\begin{equation}
\begin{array}{ll}
p_{{\rm norm},j}=&{{f_{\rm norm} }\over{\sqrt{2\pi \left[ \sigma_{\rm norm}^2+\sigma(\log N_{GC})_j^2\right]}}}\\
 & \exp\left[-{1\over 2}{{(\langle \log N_{GC}\rangle_j - \log N_{GC,0}(M^*_j))^2}\over{\sigma_{\rm norm}^2+\sigma(\log N_{GC})_j^2}}\right]\ ,
\end{array}
\end{equation}
and a probability of having an `enhanced' GCS of
\begin{equation}
\begin{array}{ll}
p_{{\rm enh},j}=& {{f_{\rm enh}}\over{\sqrt{2\pi \left[ \sigma_{\rm enh}^2+\sigma(\log N_{GC})_j^2\right]}}}\times \\
 & \exp\left[-{1\over 2}{{(\langle \log N_{GC}\rangle_j - \log N_{GC,0}(M^*_j) - \log X )^2}\over{\sigma_{\rm enh}^2+\sigma(\log N_{GC})_j^2}}\right] \ ,
\end{array}
\end{equation}
where $f_{\rm enh}=1-f_{\rm norm}$.
The parameters of this model are:
\begin{itemize}
\item{the fraction of normal galaxies, $f_{\rm norm}$;}
\item{the logarithmic shift between the two abundance relations $\log X$;}
\item{the intrinsic spreads of the two relations between stellar mass and GC abundance, for normal galaxies $\sigma_{\rm norm}$, and 
enhanced galaxies, $\sigma_{\rm enh}$.}
\end{itemize}
As the population of normal galaxies should describe the sample of G10, we impose a gaussian prior on $\sigma_{\rm norm}$, 
with properties $\sigma_{\rm norm}=0.40\pm0.05$.
The likelihood of this model is 
\begin{equation}
\mathcal{L} = \prod_j^{54} (p_{{\rm norm},j}+p_{{\rm enh},j})\ .
\end{equation} 

Our 54 measured GC abundances suggest that a model with two distinct populations is preferred: the inferred mean for the parameter
$\langle f_{\rm norm}\rangle=0.7$ is about 3-sigma away from $f_{\rm norm}=1$, which identifies a model in which all galaxies are `normal'.
At the same time, the factor $X$ is estimated at $\langle X\rangle=9.2\pm1.9$, different from 1 with a similar significance. 
This shows that some galaxies in our sample are significant outliers from the relation~(\ref{normal}).
However, this does not necessarily imply that two distinct populations are needed to describe our sample. In fact, 
we find that a model featuring a single cluster population with common properties can describe our sample equally well.
This requires a mean relation that is just slightly different than what prescribed by~(\ref{normal}), and a 
wider intrinsic spread. The main cause for this is the large relative uncertainty on the
majority of our inferred GC abundances.

The relation between stellar mass and the GCS richness of the enhanced galaxies is displayed 
by a blue dashed line in Fig.~\ref{stmass}. Although data for DF44, DFX1, DF17 and the displayed Virgo UDG were 
not used in measuring the shift $X$, the resulting relation appears to provide a reasonable fit to their properties. 
In turn, despite its extended size, DF42 may in fact be best described by the relation for normal galaxies. 
Among our 54 low surface brightness galaxies, 9 are found to 
have a probability higher than 70\% of belonging to the population of rich systems. These are highlighted
with small arrows in Fig.~\ref{stmass} and have their ID numbers shown. In fact, these are the same as the
9 systems having a GCS at least 3 times richer than expected at 90\% probability. As commented in Section~5,
not all of these galaxies are extended; in fact the majority have $R_S<1.5$~kpc. 
The physical mechanisms that make dwarf galaxies especially extended do not appear to be closely related to those that make some of
them rich in GCs. 
Deeper datasets allowing better constraints for a larger set of both normal and low surface brightness dwarfs 
(inside and outside clusters) are necessary to unequivocally determine whether cluster galaxies with 
rich GCSs are indeed a separate galaxy population.
This will help understand how GC abundance relates to galaxy size, stellar mass and morphology of galaxies
and guide the identification of the mechanisms that are responsible for these properties.

\section*{Acknowledgements}
NA and AM acknowledge stimulating discussions with Chervin Laporte. 
NA and AM thank Mike Beasley and Abraham Loeb for comments on an early version of this draft and 
are delighted to thank Pieter van Dokkum and Roberto Abraham for their useful insight during the development 
of this project and for sharing their results prior to publication.

\appendix

\section{Test suites}

Here we test our statistical framework, using purposely generated mock UDG datasets. 
We wish to evaluate our method's performance when applied to data with the properties of the CCTp data,
especially with reference to the high background counts.
Additionally, we wish to determine the prior distributions that are most appropriate for this study,
exploring any biases they might give rise to. 

\subsection{Mock datasets}
For each of our 54 UDGs, we generate a set of 100 mock datasets with the same 
footprint as the real data, and therefore the same spatial selection function. 
Mock datasets comprise a population of member GCs, centred on the UDG, and a population of contaminants, which is distributed uniformly.
For each UDG $j$, the contaminants' surface density $\Sigma_{c,j}$ is estimated using the real data, 
assuming that all catalogue sources are contaminants. This is an overestimate, but given that source counts
are indeed largely dominated by the contaminants (see Fig.~\ref{mosaic}), its relative error remains negligible.
For these tests, we do not distinguish between contaminants that appear as point sources and 
those that are extended, and are mainly concerned as to whether the richness of the GCS can be 
correctly inferred despite the strong background counts. We however seek to test whether source magnitudes
can be used to help disentangle members. Therefore, a magnitude value for each contaminant source 
is generated by sampling randomly from the magnitude distribution of the H10 catalogue, $g_{\rm H10}$, using again the fact 
that member GCs are only a very small fraction.  

The mock member GCs are generated using the following model.
\begin{itemize}
\item{The total richness of the GCS, $N_{GC}$, is generated assuming the galaxy is `normal', meaning that
it complies with the relation~(\ref{normal}). For each of the 100 mock datasets, the value of $N_{GC}$ is therefore 
sampled from a Poisson distribution around the mean value prescribed by the relation~(\ref{normal}) and the UDG's
measured stellar mass (adopted values are listed in Table~1).}
\item{The GCS has a Plummer or an exponential spatial distribution, each with a probability $p=0.5$. 
In our inference we assume a Plummer distribution: this allows us to test 
how the uncertainty on the actual profile of the UDG GCS affects our results. For either Plummer or exponential profiles, 
the half-count radius $R_{\rm h}$ has a Gaussian distribution with mean $R_{\rm h}/R_S=1.8$ and a scatter of $0.3$. 
This encompasses the properties of normal galaxies \citep[e.g.,][]{Ka14,NC16}, as well as of 
UDGs studied so far \citep{MB16a,Pe16, vD16}. }
\item{A magnitude value for each member GC is generated assuming the GCS has a Gaussian luminosity function,
with values typical for normal dwarf galaxies \citep[G10,][]{Mi07,Pe09}. Different mock datasets, however, have different 
GCLF: individual turnovers and spreads are centred respectively in $\langle F814W\rangle=27.33$~mag and 
$\sigma_{F814W}=1.1$~mag, and have a Gaussian distribution around these values with a scatter of 0.1~mag.
Our completeness correction, eqn~(\ref{incompl}), assumes $\langle F814W\rangle=27.33$ and $\sigma_{F814W}=1.1$.
As for the density profile of the GCS, this allows us to explore the effect that the uncertainty on the UDG GCLF
has on our measurements.}
\end{itemize} 

The populations of both contaminants and members are filtered by the same spatial and magnitude selections
that characterise the CCTp data. For each UDG, spatial selection excises all those mock GCCs that fall outside 
the available footprint, or in excised areas. Magnitude selection excises with probability $p=1-S_{814}(m)$ mock 
candidates with $F814W$ magnitude $m$.

\begin{figure}
\centering
\includegraphics[width=.9\columnwidth]{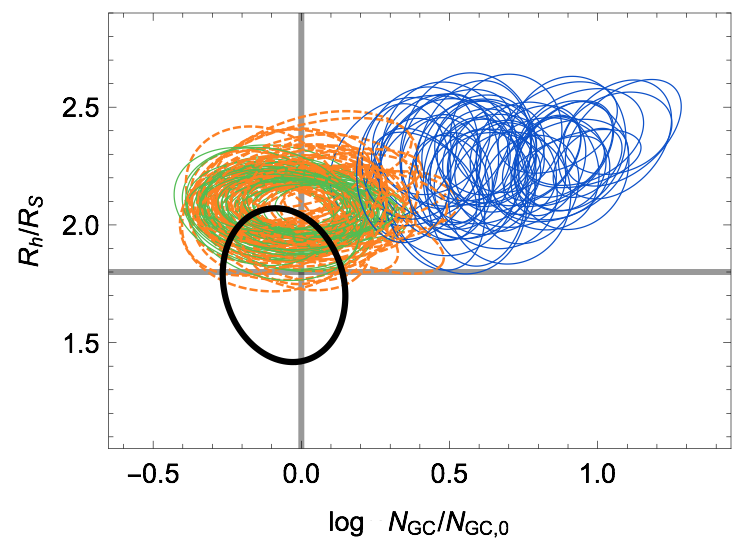}
\caption{Performance of our statistical framework. Horizontal and vertical lines indicate the
mean input parameters used to generate the mock datasets (a set of 100 for each UDG). 
For each UDG, coloured ellipses show the 1-sigma regions for the distributions of measurements 
on mock data (median GC abundance $N_{GC}$, median half-count radius $R_{\rm h}/R_S$).
The mismatch between ellipses and  lines quantifies bias.
Solid ellipses correspond to results obtained without using the source magnitudes: in blue when adopting a prior
with uniform density in $f$ and $R_h/R_S$, in green if the prior has uniform density in $\log f$ and $R_h/R_S$,
in black for our final priors, with uniform density in $\log f$ and $\log R_h/R_S$. Orange dashed ellipses 
refer to results obtained using the source magnitudes (prior with uniform density in $\log f$ and $R_h/R_S$). }
\label{bias}
\end{figure}

\subsection{Prior distribution for the fraction $f$}

As already mentioned, the main difficulty presented by the data at hand lies in the dominant background counts.
This implies that use of prior distributions that weigh differently the parameter volume $0<f_{GC}<1$ can influence our inferences.
We isolate this problem here by testing a simplified mixture model that only uses the 
spatial distribution to disentangle members and contaminants:
\begin{equation}
\mathcal{L}_j = \prod_i^{N_j} \left(p_{GC,i} + p_{{\rm cont},i}\right)\ ,
\end{equation} 
where
\begin{equation}
p_{GC,i}= f_{GC,j}{{S_{{\rm sp},j}({\bf r_i})\ \Sigma_j(r_i)}\over{\int S_{{\rm sp},j}\Sigma_j(r)}}\ ,
\end{equation} 
\begin{equation}
p_{{\rm cont},i}=(1-f_{GC,j}){{S_{{\rm sp},j}({\bf r_i})}\over{\int S_{{\rm sp},j}}}\ ,
\end{equation} 
and experiment with different prior distributions for the dimensionless free parameter $f_{GC,j}$. 
{In these tests we put ourselves in the disadvantageous position of not using the colours or 
the concentration of our test sources. This reduces our power to disentangle members GCs from contaminants,
which are differentiated by their spatial distribution alone. As a consequence, here we intend that the contaminant population
include both resolved background galaxies and the ICGC population.}

For instance, we consider a uniform prior in $f$, $0<f<1$, and a uniform prior in $\log f$. In the latter case, a finite lower bound is needed,
and we use $\log f_0 -1.5 <\log f<0$, where the value $f_0$ is the fraction of members expected in the data 
when $N_{GC}=N_{GC,0}$, i.e. if the GCS is `normal', as in eqn.~(\ref{normal}).

We first test the performance of the two different prior distributions, on all our 54 UDGs, using 100 mock datasets for each them. 
For each UDG mock dataset, we record results for the median values of the posterior 
distributions for both $R_{\rm h}/R_S$ and $N_{GC}$, after correcting the latter for completeness as in eqn~(\ref{incompl}). 
Hence, each mock dataset defines a point in this plane.  Figure~\ref{bias} displays the 1-sigma ellipse for the collection of these 
100 points, one ellipse for each of our UDGs. Blue ellipses use the prior distribution with  uniform density in $f$, 
green ellipses refer to the prior distribution with uniform density in $\log f$. The horizontal and vertical grey lines display 
the mean values of the input parameters used to generate the mock datasets: any systematic displacement 
of the ellipses quantifies bias. It is clear that the prior with uniform density in $f$ leads to a significant bias in 
both free parameters: GC abundances are significantly overestimated, by a factor $\gtrsim3$.
This is significantly ameliorated by adopting the prior distribution with uniform density in $\log f$,
for which all 54 ellipses correctly include the input value for $N_{GC}$ within 1-sigma.   

Some residual bias towards higher values is still present in our inference for the ratio $R_{\rm h}/R_S$,
for which these tests assume a uniform prior $0.5<R_{\rm h}/R_S<3.5$. To correct for this bias,
similarly to what done for the fraction $f$, we adopt a prior that is uniformly distributed in $\log R_{\rm h}/R_S$,
in the interval $0.75<R_{\rm h}/R_S<3.5$. The black ellipse in Fig.~A1 shows that this choice is appropriate.

\subsection{Using the sources magnitude}

We also wish to test whether the additional magnitude information of each source might help disentangle 
members from contaminants. To do so, we update the model at eqns.~(A2,A3) as follows:
\begin{equation}
p'_{GC,i}=p_{GC,i}\times {{\mathcal{G}_{\rm GCLF}(m_i)S_{814}(m_i)}\over{\int \mathcal{G}_{\rm GCLF}S_{814}}}\ ,
\end{equation} 
\begin{equation}
p'_{{\rm cont},i}=p_{{\rm cont},i}\times g_{\rm H10}(m_i)\ ,
\end{equation} 
where, $\mathcal{G}_{\rm GCLF}$ is the same GCLF adopted in producing the mock datasets and in performing
our incompleteness correction~(\ref{incompl}), $S_{814}$ is the completeness function described in Section~2.1,
while $g_{\rm H10}$ is the magnitude distribution of the contaminants obtained as in~A1, normalized to $\int g_{\rm H10}=1$.

Results obtained using this model (and the prior with uniform density in $\log f$) are shown by orange 
dashed ellipses in Fig.~\ref{bias}. These do not show systematic improvement with 
respect to the simpler model of eqns.~(A2,A3): the magnitude distributions of member GCs and contaminants 
are not sufficiently different from each other to guarantee a measurable statistical improvement.
As a consequence, we do not use explicitly the magnitude of each source in our analysis on the real data.

\end{document}